\documentclass[aps,prb,superscriptaddress,twocolumn]{revtex4-2}
\usepackage{amsmath}
\usepackage{amsfonts}
\usepackage{amssymb}

\usepackage{ifthen}
\usepackage[usenames]{color}
\usepackage{amssymb,mathbbol,amsmath,amsbsy}
\usepackage{amsfonts}
\usepackage{dsfont}
\usepackage{amsmath}
\usepackage{graphicx}
\usepackage[utf8]{inputenc}
\usepackage{lipsum}  
\usepackage{float}
\usepackage{xcolor}
\usepackage{soul}

\newcommand{\lsp}{\lambda_\mathrm{sp}}
\newcommand{\lmfp}{\lambda}
\newcommand{\trans}{\mathcal{T}}
\newcommand{\Nch}{N_\mathrm{ch}}
\newcommand{\supplement}{Supplemental Material}

\DeclareUnicodeCharacter{0301}{\'{e}}

\begin{document}

\author{Hazan \"Ozkan}
\affiliation{%
Department of Photonics, Izmir Institute of Technology, 35430 Urla, Izmir, Turkey.
}

\author{Stephan Roche}
\affiliation{%
	Catalan Institute of Nanoscience and Nanotechnology, CSIC and The Barcelona Institute of Science and Technology, Campus UAB, 08193 Bellaterra, Barcelona, Spain.
}
\affiliation{
	 ICREA, Institució Catalana de Recerca i Estudis Avançats, 08070 Barcelona, Spain.
}

\author{H\^{a}ldun~Sevin\c{c}li}
\email{sevincli@fen.bilkent.edu.tr}
\affiliation{%
	Department of Physics, Bilkent University, 06800 Ankara, Turkey.
}

\newcommand{\baslik}{Anomalous diffusion in multichannel systems without a Lévy distribution of disorder}

\title{\baslik}

\begin{abstract}
	We show that multichannel quantum systems with uncorrelated but asymmetric Anderson-type disorder can exhibit anomalous diffusion, even in the absence of heavy-tailed disorder. Using a minimal two-channel model with channel asymmetry, we demonstrate a  crossover from normal to anomalous transport tuned by interchannel coupling. Applied to quasi-one-dimensional lattices with edge disorder, this leads to long-tailed transmission statistics characterized by ballistic segments interspersed with localized ones, reminiscent of Lévy flights. This channel-asymmetric anomalous diffusion (CAAD) emerges from quantum interference between channels with differing disorder strengths. While CAAD governs transport at intermediate lengths, conventional localization prevails asymptotically, violating the Thouless relation. These results highlight a distinct quantum mechanism for anomalous diffusion beyond classical paradigms.
\end{abstract}

\maketitle

\clearpage

Normal diffusion is the predominant transport regime in both classical and quantum systems, characterized by Gaussian spreading and a linear scaling of the mean squared displacement with time. In contrast, anomalous diffusion arises under specific conditions and exhibits deviations from this behavior, often marked by non-Gaussian statistics and non-linear time scaling. A well-established theoretical framework accounts for anomalous diffusion through Lévy-type statistics, where the distribution of scatterers or step lengths follows a heavy-tailed (power-law) form. Such distributions give rise to rare but large displacements, fundamentally altering transport dynamics and leading to long-tailed spatial profiles.~\cite{uchaikin2011chance,feller1991introduction}
Such behavior has been observed across diverse systems,
ranging from human and animal behavior 
to light propagation and charge transport in mesoscopic systems.~\cite{brockmann2006scaling,baronchelli2013levy,sims_scaling_2008,de_jager_levy_2011,barthelemy2008levy,leadbeater1998levy,groth2012transmission,fonseca2024levy}
In single-channel quantum wires with Lévy-distributed scatterers, Beenakker \textit{et al.} showed that transmission scales non-algebraically with system length.~\cite{beenakker:prb:2009}

Anomalous diffusion is typically associated with a distinct dependence of the transmission amplitude, $\trans$, on system length $L$. This dependence can be described by a generalized diffusion equation (GDE) as
\begin{align}
	\trans(L) = \frac{\trans_0}{1 + \left( \frac{L}{\lsp} \right)^{\alpha/2}},
	\label{eqn:GDE}
\end{align}
where  $\lsp$ is the mean spread length, and
$\trans_0$ is the transmission amplitude without any scatterings.
$\trans_0$ corresponds to the number of channels, $\Nch$, and the ratio $\trans/\Nch$ represents the average transmission probability.
The diffusion exponent $\alpha$ is a dimensionless parameter characterizing the degree of anomalous diffusion.~\cite{barthelemy2008levy} For normal diffusion, $\alpha = 2$ and $\lsp$ corresponds to the mean free path ($\lmfp$), whereas for anomalous diffusion, $\alpha < 2$, indicating superdiffusive behavior.


In this Letter, we introduce a minimal yet general theoretical model that captures the essential conditions for the emergence of channel-asymmetric anomalous diffusion (CAAD) in multichannel quantum systems. Unlike conventional frameworks for anomalous diffusion, which typically rely on heavy-tailed disorder or correlated randomness, our approach reveals that CAAD can arise solely from asymmetric disorder strengths across otherwise uncorrelated channels. This mechanism uncovers a previously overlooked route to anomalous transport, rooted in quantum interference between coexisting localized and ballistic modes.
We then apply this framework to quasi-one-dimensional lattices with experimentally relevant edge disorder, demonstrating that the phenomenology predicted by the two-channel model (namely, the Lévy-like transport behavior and long-tailed transmission distributions) persists in realistic geometries. Our findings not only establish the generality of CAAD but also provide concrete criteria for its observation in practical systems, such as nanowires, topological edge states, and engineered heterostructures. Finally, we discuss experimental signatures and the implications for quantum transport control and disorder engineering.

\begin{figure}[t]
	\includegraphics[width=0.425\textwidth]
	{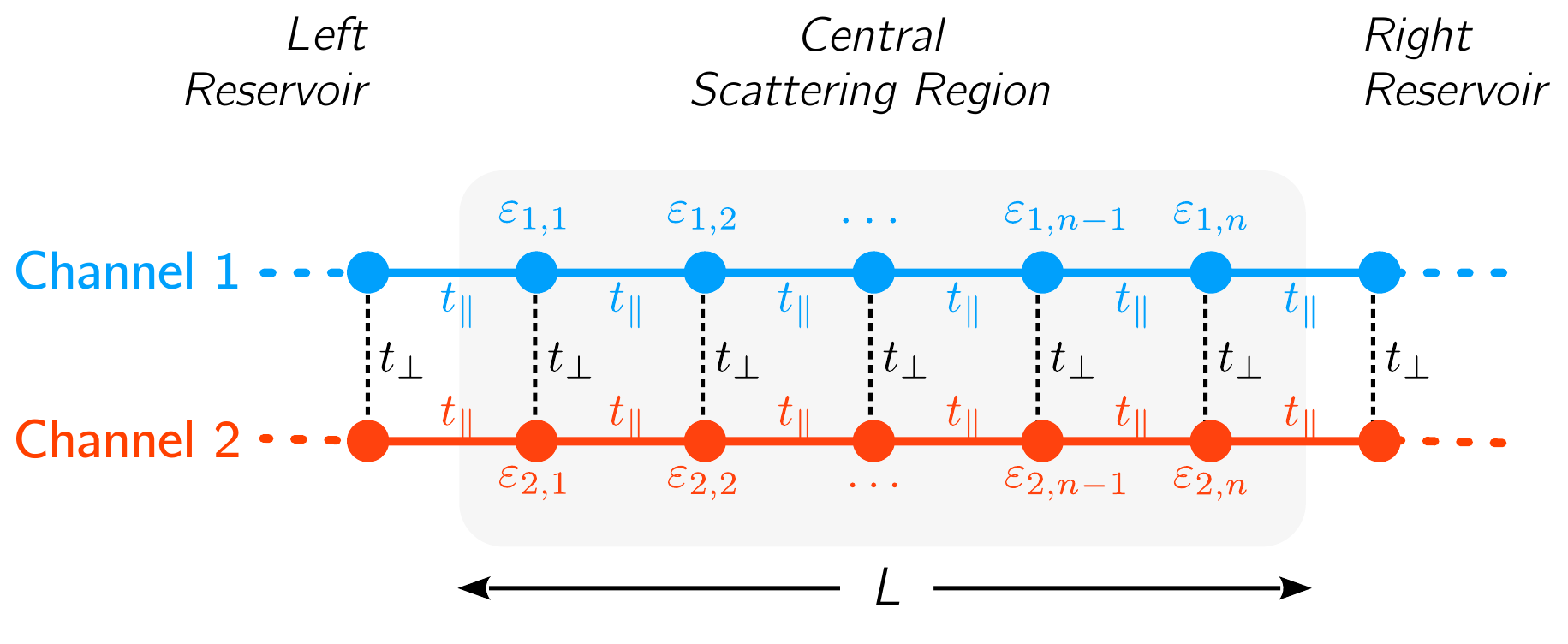}
	\vspace{-2mm}
	\includegraphics[width=0.425\textwidth]{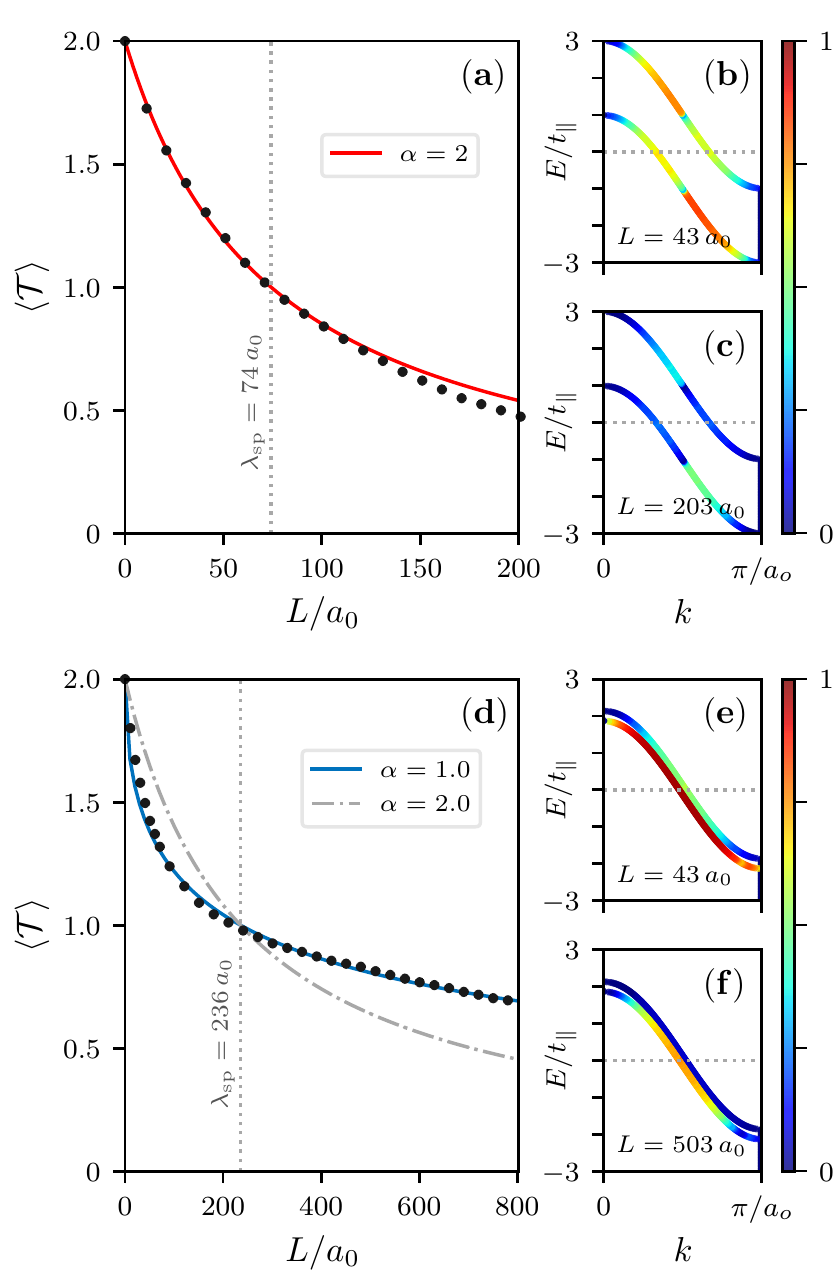}
	\vspace{-0mm}
	\caption{
		\textbf{Two-channel model of CAAD}
		demonstrating transition from normal to anomalous diffusion by varying $t_\perp/t_\parallel$.
		Length dependent transmission averaged over 1000 configurations is plotted for and $E/t_\parallel=0$ (a and d). 
		Strong interchannel coupling ($t_\perp/t_\parallel{=}1$) gives rise normal diffusion (a).  
		The contribution of each channel to transmission is resolved for each energy on the energy-band diagram with colors referring to the channel resolved transmission probabilities (b--c, colorbar is shown on the left).
		The same quantities are plotted for weak interchannel coupling ($t_{\perp}/ t_{\parallel}{=}1/20$) and shown in the same order in (d--f). 
		The comparison of transmission in normal and anomalous cases
		can be observed in (d). The contrast in channel-resolved transmission probabilities is striking as shown in panels (e) and (f).
	}
	\label{fig:fig_2channel_main}
\end{figure}


\textit{Two-Channel Model}--- We propose a minimal two-channel model to demonstrate CAAD in systems with Anderson-type disorder.
As illustrated in the top panel of Fig.~\ref{fig:fig_2channel_main}, the system consists of two linear chains with nearest-neighbor hopping $t_\parallel$. 
The Hamiltonian is
\begin{eqnarray}
	H =&
	\displaystyle\sum\limits_{\nu,j\in C} \varepsilon_{\nu j} c_{\nu j}^\dagger c_{\nu j} 
	+ t_\parallel \displaystyle\sum\limits_{\nu j} 
	\left(c_{\nu j+1}^\dagger c_{\nu j} + c_{\nu j}^\dagger c_{\nu j+1} \right) \nonumber\\
	&+ t_\perp \displaystyle\sum\limits_{j} 
	\left(c_{\nu j}^\dagger c_{\nu' j} + c_{\nu' j}^\dagger c_{\nu j} \right),
\end{eqnarray}
where $\nu$ and $j$ denote channel and site indices, respectively, and $c_{\nu j}$ ($c_{\nu j}^\dagger$) are electron annihilation (creation) operators.
The central scattering region is connected to semi-infinite reservoirs from left and right.
Interchain coupling in the transverse direction ($t_\perp$) is included, though its extension to the reservoirs is not essential (see \supplement, Section~\ref{sec:Supporitng_AlternativeGeometry}).
The chains differ only in disorder strength, which is applied exclusively within the scattering region.
Each channel hosts Anderson-type disorder, $\varepsilon_{\nu j}$.
In Channel-1, onsite energies $\varepsilon_{1j}$ are uniformly distributed in $[-t_\parallel / 2, t_\parallel / 2]$, while Channel-2 experiences weaker disorder with $\varepsilon_{2j} \in [-t_\parallel / 20, t_\parallel / 20]$.
The disorder in each channel is constrained to have zero mean, $\sum_j \varepsilon_{\nu j} = 0$.
The ratio of hopping strengths, $t_{\perp}/t_{\parallel}$, is crucial in determining transport.
It controls the degree of hybridization between the channels, influencing whether diffusion follows normal or anomalous behavior.
We examine two cases by varying this ratio: first, $t_{\perp}/t_{\parallel} = 1$, representing strong inter-channel coupling, and second, $t_{\perp}/t_{\parallel} = 1/20$, corresponding to weak inter-channel coupling.
In both cases, we assess the transmission dependence on system length and analyze channel-resolved transmission spectra to uncover transport mechanisms.

Landauer-B\"uttiker formalism and Green's function techniques are employed to compute quantum transmission amplitudes.~\cite{landauer_ibm_1957,buttiker_prl_1986, ryndyk_book_2015,datta_book_1997}
(methodological details are provided in the \supplement.)
Results for strong interchannel coupling are shown in Fig.~\ref{fig:fig_2channel_main}(a-c).
Simulation data are fitted to the GDE using $\lsp = 74\,a_0$ as the length where $\trans$ halves from its pristine value, $\trans_0$, $a_0$ being the interatomic distance.
The fitted curve (red) closely follows the normal diffusion curve (gray, dot-dashed, not visible), with diffusion exponent $\alpha = 2.0$, indicating normal diffusion.
For lengths $L > 2\lsp$, data fall below the normal diffusion curve, signaling localization onset.
Transmission data for longer systems and detailed localization analysis are discussed later.

In Fig.~\ref{fig:fig_2channel_main}(b-c), channel-resolved transmission values for strong coupling are plotted.
Values are color-coded within energy dispersion curves at short ($L \ll \lsp$) and long ($L \gg \lsp$) system lengths.
In both cases, contributions from both channels are nearly identical around the Fermi level, consistent with strong hybridization.

\begin{figure*}[t]
	\includegraphics[width=0.99\textwidth]{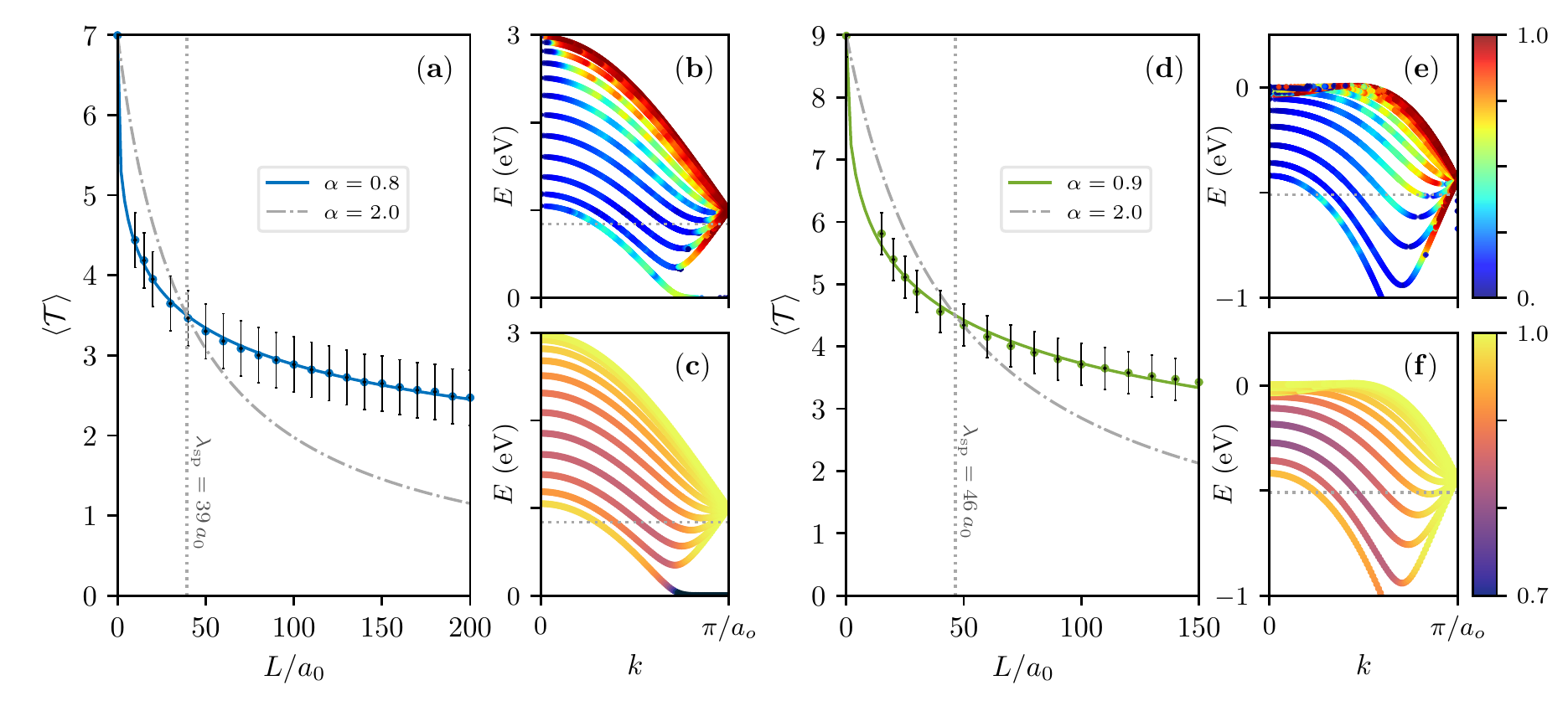}
	\caption{
			\textbf{Multichannel anomalous diffusion in quasi-one dimensional hexagonal lattices.}
			The average transmission (a,d), channel-resolved transmission probabilities (b,e), and edge/bulk character of channels (c,f) are shown for GNRs (a--c) and QNRs (d--f).
			The average transmission values depending on the length are plotted for 0.84~eV for GNR (a), and for -0.51~eV for QNR (d). 
			The band-resolved transmission probabilities (b and e) display variations for different channels.
			The contribution of the edge and bulk states to the transmission process is represented via an energy band diagram. 
		}
	\label{fig:ribbon_main}
\end{figure*}

The weak inter-channel coupling is the primary focus in this work, shown in Fig.~\ref{fig:fig_2channel_main}(d-f).
Here, $\lsp = 236\,a_0$ and the GDE fit yields $\alpha = 1.0$ (blue curve in Fig.~\ref{fig:fig_2channel_main}(d)).
Compared to normal diffusion (grey, dot-dashed), the fitted curve exhibits distinct features. At lengths shorter than $\lsp$, transmission decreases faster than normal diffusion; for $L > \lsp$, it decreases more slowly.
This contrasts with strong inter-channel coupling where localization occurs.
Such length-dependent characteristics are signatures of Lévy flights.

Channel-resolved contributions reveal the origin of anomalous behavior.
At $E = 0$ and $L {=}43\,a_0$, Channel-2 shows nearly ballistic transmission, while $\trans_1$ is about 0.5 (Fig.~\ref{fig:fig_2channel_main}(e-f)).
For $L{=}503\,a_0$, $\trans_2$ remains above 0.7, while $\trans_1$ drops below 0.1.
This indicates Channel-2 maintains quasi-ballistic transport, while Channel-1 localizes, showing coexisting transport regimes.
Strong inter-channel coupling results in unified diffusion; weak coupling shows coexisting extreme regimes.
Total transmission remains within the diffusion range, well described by the GDE (Eqn.~\ref{eqn:GDE}).

Transmission probability distributions (TPDs) provide crucial insight into CAAD.
Single-channel TPDs at ballistic, diffusive, and localized regimes guide the analysis of multichannel systems. 
For the two-channel system, TPDs at lengths 10$a_0$, 40$a_0$, 120$a_0$, 1300$a_0$, and 4000$a_0$ are examined.
Depending on length, channels show distinct ballistic, diffusive, or localized behavior similar to those in a single-channel case.
At $L=10\,a_0$, both channels are ballistic with large $\langle \trans_i \rangle$.
At $L=40\,a_0$, the unresolved TPD exhibits an unusual distribution: probable transmission values are almost evenly distributed between 1 and 2, unlike single-channel or strongly coupled two-channel systems.
Channel-resolved analysis reveal the cause: Channel-1 has a diffusive TPD, Channel-2 remains ballistic.
Consequently, total transmission is mostly $\geq1$, and the shape within $\trans \in [1,2]$ resembles the diffusive channel's distribution.
At $L=120\,a_0$, Channel-1 localizes, but Channel-2 remains ballistic.
The unresolved distribution peaks near $\trans=1$ and is asymmetric: the left side, dominated by ballistic Channel-2, has a short tail; the right side, dominated by localized Channel-1, has a broader tail.
At $L=1300\,a_0$, Channel-2 is diffusive while Channel-1 localization strengthens; unresolved TPD is roughly uniform in $[0,1]$.
At even longer lengths, only localized states contribute.
These effects are observed over a broad energy range and
the coexistence of different transport regimes and CAAD dominate  the spectrum.
(see \supplement, Sec.~\ref{sec:supp_prob_dist} for channel-resolved TPD plots)


\textit{Quasi-1D ribbons}$-$
The two-channel model predicts the onset of CAAD in certain multichannel systems when disorder strength shows substantial variations across channels. This situation can be realized in quasi-1D ribbons of two-dimensional structures with edge disorder introduced by randomly distributed vacancies at the edges.
In these systems, there exist states localized at the edges together with bulk-like states that spread across the entire ribbon width. This spatial distribution creates an imbalance in the relative strengths of disorder. Another effect of disorder is to enable inter-channel scattering.
The ribbon width plays a key role, as it determines the number of channels; wider ribbons have more channels and a lower edge-to-bulk channel ratio. Hence, ribbon width and defect concentration effectively represent the parameters $t_\perp/t_\parallel$ and the disorder strength ratio in the two-channel model.

We first study graphene nanoribbons (GNRs) with zigzag edges and 24 atoms per unit cell. Figure~\ref{fig:ribbon_main}(a) displays the average transmission at $E=0.84$~eV. As in the two-channel model, transmission values fall below the diffusion curve for $L<\lsp$ and exhibit a heavy-tailed decay for $L>\lsp$, both characteristic of anomalous diffusion. The fitted diffusion exponent is $\alpha = 0.8$.
Figure~\ref{fig:ribbon_main}(b) shows band-resolved contributions to the transmission across a wide energy range for $L=\lsp$. Figure~\ref{fig:ribbon_main}(c) presents the energy band diagram colored by edge or bulk character: yellow indicates bulk-like states, while darker colors highlight edge-dominated states. Comparing panels (b) and (c), we observe that edge states are already localized around $L \approx \lsp$, while bulk states remain quasi-ballistic.

\begin{figure}[b]
	\vspace{-5mm}
	\begin{center}
		\includegraphics[width=0.49\textwidth]{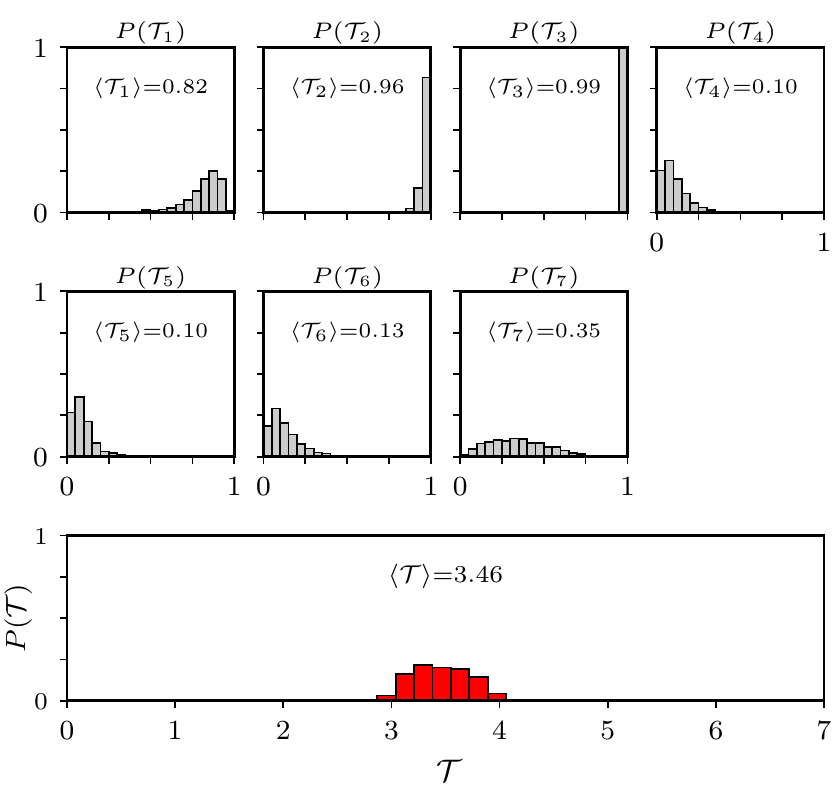}
	\end{center}
	\vspace{-5mm}
	\caption{
		\textbf{Channel-resolved probability distributions} in edge-disordered GNR for $L\sim\lsp$ reveal the same type of asymmetry in channel resolved transport regimes as in the two-channel model.
		}
	\label{fig:ProbDist_GNR}
\end{figure}

Channel-resolved TPDs of GNRs at $L{=}40\,a_0$ are shown in Fig.~\ref{fig:ProbDist_GNR}. The system contains 7 channels, and signatures of all three transport regimes are observed simultaneously: 3 channels are quasi-ballistic, 1 is diffusive, and 3 are localized. The total transmission probability distribution (TPD) exhibits a single peak around $\trans=3.4$ and mainly spans the range from 3 to 4. Despite the mixed transport regimes, $P(\trans)$ is relatively symmetric. These channel-resolved TPDs confirm that the anomalous diffusion in GNRs originates from the same mechanism as in the two-channel model.

We also examine ribbons of hexagonal lattices with quartic dispersion near their band edge (QNRs). These systems host Mexican-hat-shaped bands, leading to strong density-of-states (DOS) singularities that cause earlier localization.~\cite{sevincli2017quartic,polat_jap_2024} Furthermore, the increased channel density in QNRs enhances the emergence of CAAD.
Figure~\ref{fig:ribbon_main}(d) shows CAAD in QNRs with $\alpha=0.9$, capturing both the rapid decay at short distances and the long tail at larger lengths. The band-resolved transmission and edge/bulk characteristics in Figs.~\ref{fig:ribbon_main}(e)--(f) exhibit similar behavior to GNRs, further supporting the generality of the CAAD mechanism.
Importantly, not only hexagonal ribbon lattices but also square-lattice ribbons with edge disorder exhibit CAAD (see \supplement Sec.~\ref{sec:supp_square_lattice}).

\begin{figure}[t]
	\includegraphics[width=0.5\textwidth]{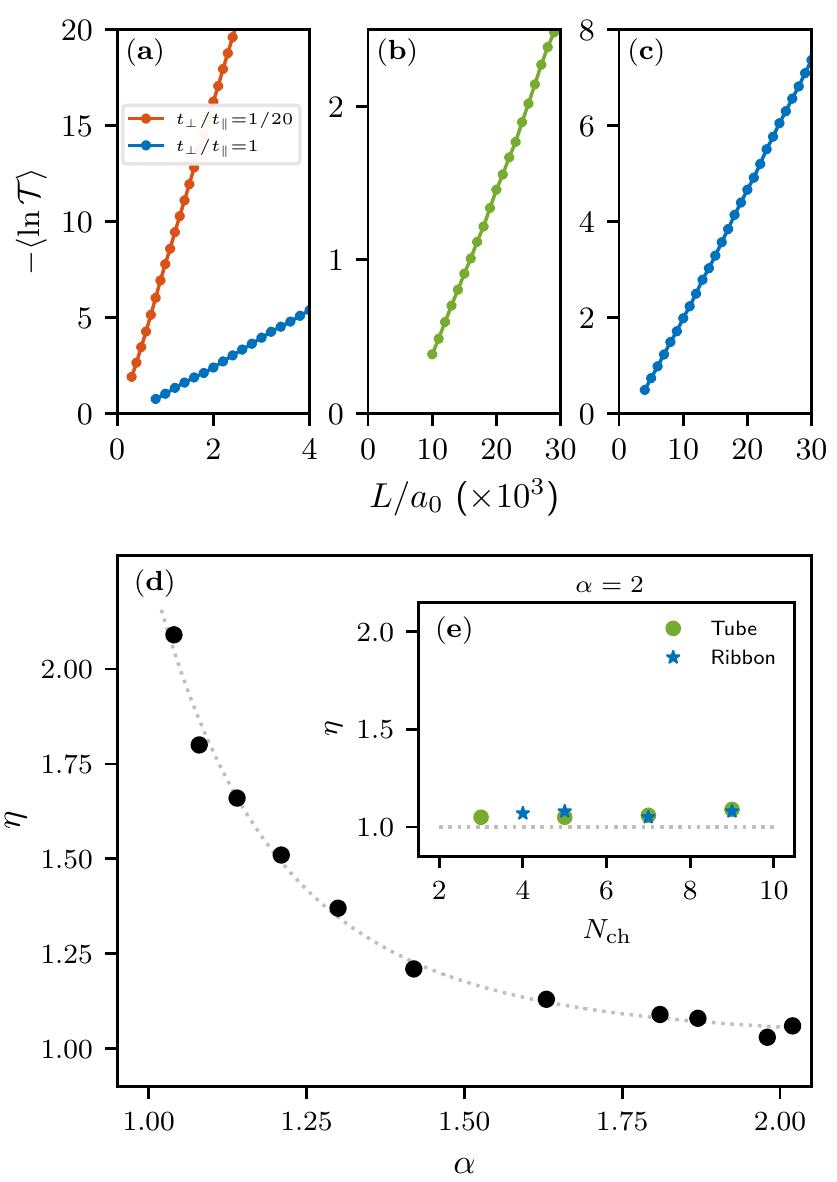}
	\caption{
		\textbf{Onset of normal localization and the Thouless relation.}
		Geometric average of the conductance depending on the system length for (a) the two-channel model with Anderson disorder, QNR and GNR with $20\%$ edge defect density (b and c). 
		The linear increase of $-\langle\ln\trans\rangle$ with system length indicates normal localization behavior.
		The validity of Thouless relation is inspected for systems showing normal and anomalous diffusion (d and e). For $\alpha=2$,  $\eta\simeq1$ is satisfied for various $\Nch$ in different geometries~(e), whereas for $\alpha<2$, $\eta$ is always larger than~1~(d).
	}
	\label{fig:localization}
\end{figure}

\textit{Localization}$-$
In quasi-one-dimensional systems, localization invariably emerges once the system length surpasses the localization length.~\cite{anderson_physrev_1958,lopez_prb_2018}
Several studies report anomalous localization in systems with Lévy-type disorder.~\cite{mendez2016lloyd,amanatidis2017coherent,fernandez2012photonic}
Here, we investigate the evolution of multichannel anomalous transport in systems much longer than the characteristic length scale $\lsp$.
The nature of localization can be analyzed using the geometric average of logarithmic transmission: a linear dependence on $L$ indicates normal localization, while a power-law dependence suggests anomalous localization.~\cite{fernandez2012photonic}
As shown in Fig.~\ref{fig:localization}, anomalous localization is not observed in any of the systems studied.
This result is consistent with our interpretation of anomalous diffusion in these systems.
At longer lengths, the coexistence of distinct transport regimes fades, and transmission probability distributions develop a single sharp peak at very low values.
Consequently, for $L \gg \lsp$, all channels are localized, and $\langle \ln \trans \rangle$ exhibits a linear dependence on $L$, characteristic of normal localization.
Although inter-channel coupling is weak at short distances—leading to poor hybridization—channels become well hybridized for $L \gg \lsp$.
In this regime, the localization length $\xi$ and the standard localization relation suffice to describe transport behavior.


\textit{Thouless Relation}$-$
The relation between length scales that define transport regimes is of central importance.
Following Anderson’s demonstration of localization in single-channel 1D systems,~\cite{anderson_physrev_1958}
Thouless extended these ideas to multichannel quasi-1D systems, assuming uniform disorder and well-hybridized channels. He proposed a relation between the mean-free-path $\lambda$ and localization length $\xi$, namely $\xi \sim \Nch \lambda$.~\cite{thouless_jphysc_1973,thouless_prl_1977}
A more refined expression, derived using random matrix theory, is given by~\cite{beenakker_rmp_1997}
\begin{align}
	\label{eqn:thouless}
	\xi = \frac{N_\mathrm{ch} + 1}{2} \lambda.
\end{align}
Having established that the systems under consideration exhibit a crossover from anomalous diffusion to normal localization at sufficiently long lengths, a natural question arises: does the Thouless relation hold between $\lsp$ and $\xi$ in these systems?

To address this, we compare our simulation results with Eqn.~\ref{eqn:thouless}.
Figure~\ref{fig:localization}(d) displays the ratio $\eta = 2\xi / \left( (N_\mathrm{ch}+1)\lsp \right)$ as a function of $\alpha$.
In the anomalous diffusion regime ($\alpha < 2$), $\eta$ exceeds 1, and increases as $\alpha$ decreases.
This deviation signals suppressed localization, consistent with the long-tail behavior observed during anomalous diffusion.
For $\alpha = 2$, the expected result $\eta = 1$ is satisfied across various $N_\mathrm{ch}$ values in both ribbon and tubular geometries, as shown in the inset.
Thus, $\eta > 1$ serves as another hallmark of CAAD in multichannel systems.

\textit{Four-Probe Resistance}$-$
The total resistance of the system can be decomposed into local and nonlocal components.
The nonlocal (contact) resistance is given by $R_0/\Nch$, where $R_0 = h/2e^2$ is the resistance quantum. The nonlocal contribution is independent of system length.
The local part is also known as the four-probe resistance, $R_4$, and its scaling with system length can be expressed as
\begin{align}
	R_4 = \frac{R_0}{\Nch} \left( \frac{L}{\lsp} \right)^{\alpha/2}.
\end{align}

In normal conductors, $R_4$ increases linearly with $L$, whereas under anomalous diffusion, the scaling becomes sublinear.
The contrasting behavior of $R_4$ as a function of $L$ between normal and anomalous conductors 
and the four-probe resistance of quasi-1D hexagonal lattices can be found in Fig.~\ref{fig:Supp_FourProbe}.
These results indicate that $R_4$ measurements at varying lengths could provide a conclusive experimental test for CAAD.
Experimentally, metallic nanowires encapsulated in CNTs~\cite{pham_science_2018,pham_prl_2020} present promising platforms to investigate CAAD in quasi-1D systems.


\textit{Conclusion}$-$
We have shown that anomalous diffusion can emerge in multichannel quantum systems without the need for Lévy-type disorder, provided that disorder strengths are strongly asymmetric across channels and inter-channel coupling is weak. Using a minimal two-channel model and quasi-one-dimensional extensions, we identified a robust mechanism, channel-asymmetric anomalous diffusion (CAAD), characterized by the coexistence of ballistic and localized transport regimes. This coexistence leads to long-tailed transmission statistics reminiscent of Lévy flights, despite the absence of heavy-tailed disorder.
By analyzing edge-disordered nanoribbons, we confirmed that CAAD persists in realistic geometries and across broad energy ranges. While CAAD dominates transport at intermediate scales, normal localization eventually takes over at longer lengths, marking a crossover between distinct regimes. These findings uncover a previously unrecognized quantum mechanism for anomalous diffusion, expand the theoretical landscape of mesoscopic transport, and suggest new strategies for disorder engineering and transport control in nanoscale devices.


\medskip
{\noindent{\textbf{Acknowledgements.}}}
This work was supported by The Scientific and Technological Research Council of Turkey (T\"UB\.ITAK, 1001 Grant, Project No.~119F353), and the Air Force Office of Scientific Research (AFOSR, Award No.~FA9550-21-1-0261).
S.R acknowledge funding from European Union ``NextGenerationEU/PRTR" under grant PCI2021-122035-2A-2a. ICN2 is funded by the CERCA Programme/Generalitat de Catalunya and supported by the Severo Ochoa Centres of Excellence programme, Grant CEX2021-001214-S, funded by MCIN/AEI/10.13039.501100011033. This work is also supported by MICIN with European funds- NextGenerationEU (PRTR-C17.I1) and by 2021 SGR 00997, funded by Generalitat de Catalunya.

\bibliographystyle{apsrev4-2}
\bibliography{biblio.bib}

\begin{thebibliography}{30}%
\makeatletter
\providecommand \@ifxundefined [1]{%
 \@ifx{#1\undefined}
}%
\providecommand \@ifnum [1]{%
 \ifnum #1\expandafter \@firstoftwo
 \else \expandafter \@secondoftwo
 \fi
}%
\providecommand \@ifx [1]{%
 \ifx #1\expandafter \@firstoftwo
 \else \expandafter \@secondoftwo
 \fi
}%
\providecommand \natexlab [1]{#1}%
\providecommand \enquote  [1]{``#1''}%
\providecommand \bibnamefont  [1]{#1}%
\providecommand \bibfnamefont [1]{#1}%
\providecommand \citenamefont [1]{#1}%
\providecommand \href@noop [0]{\@secondoftwo}%
\providecommand \href [0]{\begingroup \@sanitize@url \@href}%
\providecommand \@href[1]{\@@startlink{#1}\@@href}%
\providecommand \@@href[1]{\endgroup#1\@@endlink}%
\providecommand \@sanitize@url [0]{\catcode `\\12\catcode `\$12\catcode
  `\&12\catcode `\#12\catcode `\^12\catcode `\_12\catcode `\%12\relax}%
\providecommand \@@startlink[1]{}%
\providecommand \@@endlink[0]{}%
\providecommand \url  [0]{\begingroup\@sanitize@url \@url }%
\providecommand \@url [1]{\endgroup\@href {#1}{\urlprefix }}%
\providecommand \urlprefix  [0]{URL }%
\providecommand \Eprint [0]{\href }%
\providecommand \doibase [0]{https://doi.org/}%
\providecommand \selectlanguage [0]{\@gobble}%
\providecommand \bibinfo  [0]{\@secondoftwo}%
\providecommand \bibfield  [0]{\@secondoftwo}%
\providecommand \translation [1]{[#1]}%
\providecommand \BibitemOpen [0]{}%
\providecommand \bibitemStop [0]{}%
\providecommand \bibitemNoStop [0]{.\EOS\space}%
\providecommand \EOS [0]{\spacefactor3000\relax}%
\providecommand \BibitemShut  [1]{\csname bibitem#1\endcsname}%
\let\auto@bib@innerbib\@empty
\bibitem [{\citenamefont {Uchaikin}\ and\ \citenamefont
  {Zolotarev}(2011)}]{uchaikin2011chance}%
  \BibitemOpen
  \bibfield  {author} {\bibinfo {author} {\bibfnamefont {V.~V.}\ \bibnamefont
  {Uchaikin}}\ and\ \bibinfo {author} {\bibfnamefont {V.~M.}\ \bibnamefont
  {Zolotarev}},\ }\href@noop {} {\emph {\bibinfo {title} {Chance and stability:
  stable distributions and their applications}}}\ (\bibinfo  {publisher}
  {Walter de Gruyter},\ \bibinfo {year} {2011})\BibitemShut {NoStop}%
\bibitem [{\citenamefont {Feller}(1991)}]{feller1991introduction}%
  \BibitemOpen
  \bibfield  {author} {\bibinfo {author} {\bibfnamefont {W.}~\bibnamefont
  {Feller}},\ }\href@noop {} {\emph {\bibinfo {title} {An introduction to
  probability theory and its applications, Volume 2}}},\ Vol.~\bibinfo {volume}
  {81}\ (\bibinfo  {publisher} {John Wiley \& Sons},\ \bibinfo {year}
  {1991})\BibitemShut {NoStop}%
\bibitem [{\citenamefont {Brockmann}\ \emph {et~al.}(2006)\citenamefont
  {Brockmann}, \citenamefont {Hufnagel},\ and\ \citenamefont
  {Geisel}}]{brockmann2006scaling}%
  \BibitemOpen
  \bibfield  {author} {\bibinfo {author} {\bibfnamefont {D.}~\bibnamefont
  {Brockmann}}, \bibinfo {author} {\bibfnamefont {L.}~\bibnamefont
  {Hufnagel}},\ and\ \bibinfo {author} {\bibfnamefont {T.}~\bibnamefont
  {Geisel}},\ }\href@noop {} {\bibfield  {journal} {\bibinfo  {journal}
  {Nature}\ }\textbf {\bibinfo {volume} {439}},\ \bibinfo {pages} {462}
  (\bibinfo {year} {2006})}\BibitemShut {NoStop}%
\bibitem [{\citenamefont {Baronchelli}\ and\ \citenamefont
  {Radicchi}(2013)}]{baronchelli2013levy}%
  \BibitemOpen
  \bibfield  {author} {\bibinfo {author} {\bibfnamefont {A.}~\bibnamefont
  {Baronchelli}}\ and\ \bibinfo {author} {\bibfnamefont {F.}~\bibnamefont
  {Radicchi}},\ }\href@noop {} {\bibfield  {journal} {\bibinfo  {journal}
  {Chaos, Solitons \& Fractals}\ }\textbf {\bibinfo {volume} {56}},\ \bibinfo
  {pages} {101} (\bibinfo {year} {2013})}\BibitemShut {NoStop}%
\bibitem [{\citenamefont {Sims}\ \emph {et~al.}(2008)\citenamefont {Sims},
  \citenamefont {Southall}, \citenamefont {Humphries}, \citenamefont {Hays},
  \citenamefont {Bradshaw}, \citenamefont {Pitchford}, \citenamefont {James},
  \citenamefont {Ahmed}, \citenamefont {Brierley}, \citenamefont {Hindell},
  \citenamefont {Morritt}, \citenamefont {Musyl}, \citenamefont {Righton},
  \citenamefont {Shepard}, \citenamefont {Wearmouth}, \citenamefont {Wilson},
  \citenamefont {Witt},\ and\ \citenamefont {Metcalfe}}]{sims_scaling_2008}%
  \BibitemOpen
  \bibfield  {author} {\bibinfo {author} {\bibfnamefont {D.~W.}\ \bibnamefont
  {Sims}}, \bibinfo {author} {\bibfnamefont {E.~J.}\ \bibnamefont {Southall}},
  \bibinfo {author} {\bibfnamefont {N.~E.}\ \bibnamefont {Humphries}}, \bibinfo
  {author} {\bibfnamefont {G.~C.}\ \bibnamefont {Hays}}, \bibinfo {author}
  {\bibfnamefont {C.~J.~A.}\ \bibnamefont {Bradshaw}}, \bibinfo {author}
  {\bibfnamefont {J.~W.}\ \bibnamefont {Pitchford}}, \bibinfo {author}
  {\bibfnamefont {A.}~\bibnamefont {James}}, \bibinfo {author} {\bibfnamefont
  {M.~Z.}\ \bibnamefont {Ahmed}}, \bibinfo {author} {\bibfnamefont {A.~S.}\
  \bibnamefont {Brierley}}, \bibinfo {author} {\bibfnamefont {M.~A.}\
  \bibnamefont {Hindell}}, \bibinfo {author} {\bibfnamefont {D.}~\bibnamefont
  {Morritt}}, \bibinfo {author} {\bibfnamefont {M.~K.}\ \bibnamefont {Musyl}},
  \bibinfo {author} {\bibfnamefont {D.}~\bibnamefont {Righton}}, \bibinfo
  {author} {\bibfnamefont {E.~L.~C.}\ \bibnamefont {Shepard}}, \bibinfo
  {author} {\bibfnamefont {V.~J.}\ \bibnamefont {Wearmouth}}, \bibinfo {author}
  {\bibfnamefont {R.~P.}\ \bibnamefont {Wilson}}, \bibinfo {author}
  {\bibfnamefont {M.~J.}\ \bibnamefont {Witt}},\ and\ \bibinfo {author}
  {\bibfnamefont {J.~D.}\ \bibnamefont {Metcalfe}},\ }\href@noop {} {\bibfield
  {journal} {\bibinfo  {journal} {Nature}\ }\textbf {\bibinfo {volume} {451}},\
  \bibinfo {pages} {1098} (\bibinfo {year} {2008})}\BibitemShut {NoStop}%
\bibitem [{\citenamefont {de~Jager}\ \emph {et~al.}(2011)\citenamefont
  {de~Jager}, \citenamefont {Weissing}, \citenamefont {Herman}, \citenamefont
  {Nolet},\ and\ \citenamefont {van~de Koppel}}]{de_jager_levy_2011}%
  \BibitemOpen
  \bibfield  {author} {\bibinfo {author} {\bibfnamefont {M.}~\bibnamefont
  {de~Jager}}, \bibinfo {author} {\bibfnamefont {F.~J.}\ \bibnamefont
  {Weissing}}, \bibinfo {author} {\bibfnamefont {P.~M.~J.}\ \bibnamefont
  {Herman}}, \bibinfo {author} {\bibfnamefont {B.~A.}\ \bibnamefont {Nolet}},\
  and\ \bibinfo {author} {\bibfnamefont {J.}~\bibnamefont {van~de Koppel}},\
  }\href@noop {} {\bibfield  {journal} {\bibinfo  {journal} {Science}\ }\textbf
  {\bibinfo {volume} {332}} (\bibinfo {year} {2011})}\BibitemShut {NoStop}%
\bibitem [{\citenamefont {Barthelemy}\ \emph {et~al.}(2008)\citenamefont
  {Barthelemy}, \citenamefont {Bertolotti},\ and\ \citenamefont
  {Wiersma}}]{barthelemy2008levy}%
  \BibitemOpen
  \bibfield  {author} {\bibinfo {author} {\bibfnamefont {P.}~\bibnamefont
  {Barthelemy}}, \bibinfo {author} {\bibfnamefont {J.}~\bibnamefont
  {Bertolotti}},\ and\ \bibinfo {author} {\bibfnamefont {D.~S.}\ \bibnamefont
  {Wiersma}},\ }\href@noop {} {\bibfield  {journal} {\bibinfo  {journal}
  {Nature}\ }\textbf {\bibinfo {volume} {453}},\ \bibinfo {pages} {495}
  (\bibinfo {year} {2008})}\BibitemShut {NoStop}%
\bibitem [{\citenamefont {Leadbeater}\ \emph {et~al.}(1998)\citenamefont
  {Leadbeater}, \citenamefont {Falko},\ and\ \citenamefont
  {Lambert}}]{leadbeater1998levy}%
  \BibitemOpen
  \bibfield  {author} {\bibinfo {author} {\bibfnamefont {M.}~\bibnamefont
  {Leadbeater}}, \bibinfo {author} {\bibfnamefont {V.}~\bibnamefont {Falko}},\
  and\ \bibinfo {author} {\bibfnamefont {C.}~\bibnamefont {Lambert}},\
  }\href@noop {} {\bibfield  {journal} {\bibinfo  {journal} {Physical Review
  Letters}\ }\textbf {\bibinfo {volume} {81}},\ \bibinfo {pages} {1274}
  (\bibinfo {year} {1998})}\BibitemShut {NoStop}%
\bibitem [{\citenamefont {Groth}\ \emph {et~al.}(2012)\citenamefont {Groth},
  \citenamefont {Akhmerov},\ and\ \citenamefont
  {Beenakker}}]{groth2012transmission}%
  \BibitemOpen
  \bibfield  {author} {\bibinfo {author} {\bibfnamefont {C.}~\bibnamefont
  {Groth}}, \bibinfo {author} {\bibfnamefont {A.~R.}\ \bibnamefont
  {Akhmerov}},\ and\ \bibinfo {author} {\bibfnamefont {C.}~\bibnamefont
  {Beenakker}},\ }\href@noop {} {\bibfield  {journal} {\bibinfo  {journal}
  {Physical Review E—Statistical, Nonlinear, and Soft Matter Physics}\
  }\textbf {\bibinfo {volume} {85}},\ \bibinfo {pages} {021138} (\bibinfo
  {year} {2012})}\BibitemShut {NoStop}%
\bibitem [{\citenamefont {Fonseca}\ \emph {et~al.}(2024)\citenamefont
  {Fonseca}, \citenamefont {Barbosa},\ and\ \citenamefont
  {Pereira}}]{fonseca2024levy}%
  \BibitemOpen
  \bibfield  {author} {\bibinfo {author} {\bibfnamefont {D.~B.}\ \bibnamefont
  {Fonseca}}, \bibinfo {author} {\bibfnamefont {A.~L.}\ \bibnamefont
  {Barbosa}},\ and\ \bibinfo {author} {\bibfnamefont {L.~F.~C.}\ \bibnamefont
  {Pereira}},\ }\href@noop {} {\bibfield  {journal} {\bibinfo  {journal}
  {Physical Review B}\ }\textbf {\bibinfo {volume} {110}},\ \bibinfo {pages}
  {075421} (\bibinfo {year} {2024})}\BibitemShut {NoStop}%
\bibitem [{\citenamefont {Beenakker}\ \emph {et~al.}(2009)\citenamefont
  {Beenakker}, \citenamefont {Groth},\ and\ \citenamefont
  {Akhmerov}}]{beenakker:prb:2009}%
  \BibitemOpen
  \bibfield  {author} {\bibinfo {author} {\bibfnamefont {C.~W.~J.}\
  \bibnamefont {Beenakker}}, \bibinfo {author} {\bibfnamefont {C.~W.}\
  \bibnamefont {Groth}},\ and\ \bibinfo {author} {\bibfnamefont {A.~R.}\
  \bibnamefont {Akhmerov}},\ }\href@noop {} {\bibfield  {journal} {\bibinfo
  {journal} {Physical Review B}\ }\textbf {\bibinfo {volume} {79}},\ \bibinfo
  {pages} {024204} (\bibinfo {year} {2009})}\BibitemShut {NoStop}%
\bibitem [{\citenamefont {Landauer}(1957)}]{landauer_ibm_1957}%
  \BibitemOpen
  \bibfield  {author} {\bibinfo {author} {\bibfnamefont {R.}~\bibnamefont
  {Landauer}},\ }\href@noop {} {\bibfield  {journal} {\bibinfo  {journal} {IBM
  Journal of Research and Development}\ }\textbf {\bibinfo {volume} {1}},\
  \bibinfo {pages} {223} (\bibinfo {year} {1957})}\BibitemShut {NoStop}%
\bibitem [{\citenamefont {B\"uttiker}(1986)}]{buttiker_prl_1986}%
  \BibitemOpen
  \bibfield  {author} {\bibinfo {author} {\bibfnamefont {M.}~\bibnamefont
  {B\"uttiker}},\ }\href@noop {} {\bibfield  {journal} {\bibinfo  {journal}
  {Phys. Rev. Lett.}\ }\textbf {\bibinfo {volume} {57}},\ \bibinfo {pages}
  {1761} (\bibinfo {year} {1986})}\BibitemShut {NoStop}%
\bibitem [{\citenamefont {Ryndyk}(2015)}]{ryndyk_book_2015}%
  \BibitemOpen
  \bibfield  {author} {\bibinfo {author} {\bibfnamefont {D.}~\bibnamefont
  {Ryndyk}},\ }\href@noop {} {\emph {\bibinfo {title} {Theory of Quantum
  Transport at Nanoscale: An Introduction}}}\ (\bibinfo  {publisher}
  {Springer},\ \bibinfo {year} {2015})\BibitemShut {NoStop}%
\bibitem [{\citenamefont {Datta}(1997)}]{datta_book_1997}%
  \BibitemOpen
  \bibfield  {author} {\bibinfo {author} {\bibfnamefont {S.}~\bibnamefont
  {Datta}},\ }\href@noop {} {\emph {\bibinfo {title} {Electronic {Transport} in
  {Mesoscopic} {Systems}}}}\ (\bibinfo  {publisher} {Cambridge University
  Press},\ \bibinfo {address} {Cambridge, UK; New York},\ \bibinfo {year}
  {1997})\BibitemShut {NoStop}%
\bibitem [{\citenamefont {Sevin\c{c}li}(2017)}]{sevincli2017quartic}%
  \BibitemOpen
  \bibfield  {author} {\bibinfo {author} {\bibfnamefont {H.}~\bibnamefont
  {Sevin\c{c}li}},\ }\href@noop {} {\bibfield  {journal} {\bibinfo  {journal}
  {Nano Letters}\ }\textbf {\bibinfo {volume} {17}},\ \bibinfo {pages} {2589}
  (\bibinfo {year} {2017})}\BibitemShut {NoStop}%
\bibitem [{\citenamefont {Polat}\ \emph {et~al.}(2024)\citenamefont {Polat},
  \citenamefont {\"Ozkan},\ and\ \citenamefont
  {Sevin\c{c}li}}]{polat_jap_2024}%
  \BibitemOpen
  \bibfield  {author} {\bibinfo {author} {\bibfnamefont {M.}~\bibnamefont
  {Polat}}, \bibinfo {author} {\bibfnamefont {H.}~\bibnamefont {\"Ozkan}},\
  and\ \bibinfo {author} {\bibfnamefont {H.}~\bibnamefont {Sevin\c{c}li}},\
  }\href@noop {} {\bibfield  {journal} {\bibinfo  {journal} {Journal of Applied
  Physics}\ }\textbf {\bibinfo {volume} {135}},\ \bibinfo {pages} {164301}
  (\bibinfo {year} {2024})}\BibitemShut {NoStop}%
\bibitem [{\citenamefont {Anderson}(1958)}]{anderson_physrev_1958}%
  \BibitemOpen
  \bibfield  {author} {\bibinfo {author} {\bibfnamefont {P.~W.}\ \bibnamefont
  {Anderson}},\ }\href@noop {} {\bibfield  {journal} {\bibinfo  {journal}
  {Physical Review}\ }\textbf {\bibinfo {volume} {109}},\ \bibinfo {pages}
  {1492} (\bibinfo {year} {1958})}\BibitemShut {NoStop}%
\bibitem [{\citenamefont {Lopez-Bezanilla}\ \emph {et~al.}(2018)\citenamefont
  {Lopez-Bezanilla}, \citenamefont {Froufe-P\'erez}, \citenamefont {Roche},\
  and\ \citenamefont {S\'aenz}}]{lopez_prb_2018}%
  \BibitemOpen
  \bibfield  {author} {\bibinfo {author} {\bibfnamefont {A.}~\bibnamefont
  {Lopez-Bezanilla}}, \bibinfo {author} {\bibfnamefont {L.~S.}\ \bibnamefont
  {Froufe-P\'erez}}, \bibinfo {author} {\bibfnamefont {S.}~\bibnamefont
  {Roche}},\ and\ \bibinfo {author} {\bibfnamefont {J.~J.}\ \bibnamefont
  {S\'aenz}},\ }\href {https://doi.org/10.1103/PhysRevB.98.235423} {\bibfield
  {journal} {\bibinfo  {journal} {Phys. Rev. B}\ }\textbf {\bibinfo {volume}
  {98}},\ \bibinfo {pages} {235423} (\bibinfo {year} {2018})}\BibitemShut
  {NoStop}%
\bibitem [{\citenamefont {M{\'e}ndez-Berm{\'u}dez}\ \emph
  {et~al.}(2016)\citenamefont {M{\'e}ndez-Berm{\'u}dez}, \citenamefont
  {Mart{\'\i}nez-Mendoza}, \citenamefont {Gopar},\ and\ \citenamefont
  {Varga}}]{mendez2016lloyd}%
  \BibitemOpen
  \bibfield  {author} {\bibinfo {author} {\bibfnamefont {J.}~\bibnamefont
  {M{\'e}ndez-Berm{\'u}dez}}, \bibinfo {author} {\bibfnamefont
  {A.}~\bibnamefont {Mart{\'\i}nez-Mendoza}}, \bibinfo {author} {\bibfnamefont
  {V.}~\bibnamefont {Gopar}},\ and\ \bibinfo {author} {\bibfnamefont
  {I.}~\bibnamefont {Varga}},\ }\href@noop {} {\bibfield  {journal} {\bibinfo
  {journal} {Physical Review E}\ }\textbf {\bibinfo {volume} {93}},\ \bibinfo
  {pages} {012135} (\bibinfo {year} {2016})}\BibitemShut {NoStop}%
\bibitem [{\citenamefont {Amanatidis}\ \emph {et~al.}(2017)\citenamefont
  {Amanatidis}, \citenamefont {Kleftogiannis}, \citenamefont {Falceto},\ and\
  \citenamefont {Gopar}}]{amanatidis2017coherent}%
  \BibitemOpen
  \bibfield  {author} {\bibinfo {author} {\bibfnamefont {I.}~\bibnamefont
  {Amanatidis}}, \bibinfo {author} {\bibfnamefont {I.}~\bibnamefont
  {Kleftogiannis}}, \bibinfo {author} {\bibfnamefont {F.}~\bibnamefont
  {Falceto}},\ and\ \bibinfo {author} {\bibfnamefont {V.~A.}\ \bibnamefont
  {Gopar}},\ }\href@noop {} {\bibfield  {journal} {\bibinfo  {journal}
  {Physical Review E}\ }\textbf {\bibinfo {volume} {96}},\ \bibinfo {pages}
  {062141} (\bibinfo {year} {2017})}\BibitemShut {NoStop}%
\bibitem [{\citenamefont {Fern{\'a}ndez-Mar{\'\i}n}\ \emph
  {et~al.}(2012)\citenamefont {Fern{\'a}ndez-Mar{\'\i}n}, \citenamefont
  {M{\'e}ndez-Berm{\'u}dez},\ and\ \citenamefont
  {Gopar}}]{fernandez2012photonic}%
  \BibitemOpen
  \bibfield  {author} {\bibinfo {author} {\bibfnamefont {A.}~\bibnamefont
  {Fern{\'a}ndez-Mar{\'\i}n}}, \bibinfo {author} {\bibfnamefont
  {J.}~\bibnamefont {M{\'e}ndez-Berm{\'u}dez}},\ and\ \bibinfo {author}
  {\bibfnamefont {V.~A.}\ \bibnamefont {Gopar}},\ }\href@noop {} {\bibfield
  {journal} {\bibinfo  {journal} {Physical Review A}\ }\textbf {\bibinfo
  {volume} {85}},\ \bibinfo {pages} {035803} (\bibinfo {year}
  {2012})}\BibitemShut {NoStop}%
\bibitem [{\citenamefont {Thouless}(1973)}]{thouless_jphysc_1973}%
  \BibitemOpen
  \bibfield  {author} {\bibinfo {author} {\bibfnamefont {D.~J.}\ \bibnamefont
  {Thouless}},\ }\href@noop {} {\bibfield  {journal} {\bibinfo  {journal}
  {Journal of Physics C: Solid State Physics}\ }\textbf {\bibinfo {volume}
  {6}},\ \bibinfo {pages} {L49} (\bibinfo {year} {1973})}\BibitemShut {NoStop}%
\bibitem [{\citenamefont {Thouless}(1977)}]{thouless_prl_1977}%
  \BibitemOpen
  \bibfield  {author} {\bibinfo {author} {\bibfnamefont {D.~J.}\ \bibnamefont
  {Thouless}},\ }\href@noop {} {\bibfield  {journal} {\bibinfo  {journal}
  {Physical Review Letters}\ }\textbf {\bibinfo {volume} {39}},\ \bibinfo
  {pages} {1167} (\bibinfo {year} {1977})}\BibitemShut {NoStop}%
\bibitem [{\citenamefont {Beenakker}(1997)}]{beenakker_rmp_1997}%
  \BibitemOpen
  \bibfield  {author} {\bibinfo {author} {\bibfnamefont {C.~W.~J.}\
  \bibnamefont {Beenakker}},\ }\href@noop {} {\bibfield  {journal} {\bibinfo
  {journal} {Reviews of Modern Physics}\ }\textbf {\bibinfo {volume} {69}},\
  \bibinfo {pages} {731} (\bibinfo {year} {1997})}\BibitemShut {NoStop}%
\bibitem [{\citenamefont {Pham}\ \emph {et~al.}(2018)\citenamefont {Pham},
  \citenamefont {Oh}, \citenamefont {Stetz}, \citenamefont {Onishi},
  \citenamefont {Kisielowski}, \citenamefont {Cohen},\ and\ \citenamefont
  {Zettl}}]{pham_science_2018}%
  \BibitemOpen
  \bibfield  {author} {\bibinfo {author} {\bibfnamefont {T.}~\bibnamefont
  {Pham}}, \bibinfo {author} {\bibfnamefont {S.}~\bibnamefont {Oh}}, \bibinfo
  {author} {\bibfnamefont {P.}~\bibnamefont {Stetz}}, \bibinfo {author}
  {\bibfnamefont {S.}~\bibnamefont {Onishi}}, \bibinfo {author} {\bibfnamefont
  {C.}~\bibnamefont {Kisielowski}}, \bibinfo {author} {\bibfnamefont {M.~L.}\
  \bibnamefont {Cohen}},\ and\ \bibinfo {author} {\bibfnamefont
  {A.}~\bibnamefont {Zettl}},\ }\href@noop {} {\bibfield  {journal} {\bibinfo
  {journal} {Science}\ }\textbf {\bibinfo {volume} {361}},\ \bibinfo {pages}
  {263} (\bibinfo {year} {2018})}\BibitemShut {NoStop}%
\bibitem [{\citenamefont {Pham}\ \emph {et~al.}(2020)\citenamefont {Pham},
  \citenamefont {Oh}, \citenamefont {Stonemeyer}, \citenamefont {Shevitski},
  \citenamefont {Cain}, \citenamefont {Song}, \citenamefont {Ercius},
  \citenamefont {Cohen},\ and\ \citenamefont {Zettl}}]{pham_prl_2020}%
  \BibitemOpen
  \bibfield  {author} {\bibinfo {author} {\bibfnamefont {T.}~\bibnamefont
  {Pham}}, \bibinfo {author} {\bibfnamefont {S.}~\bibnamefont {Oh}}, \bibinfo
  {author} {\bibfnamefont {S.}~\bibnamefont {Stonemeyer}}, \bibinfo {author}
  {\bibfnamefont {B.}~\bibnamefont {Shevitski}}, \bibinfo {author}
  {\bibfnamefont {J.~D.}\ \bibnamefont {Cain}}, \bibinfo {author}
  {\bibfnamefont {C.}~\bibnamefont {Song}}, \bibinfo {author} {\bibfnamefont
  {P.}~\bibnamefont {Ercius}}, \bibinfo {author} {\bibfnamefont {M.~L.}\
  \bibnamefont {Cohen}},\ and\ \bibinfo {author} {\bibfnamefont
  {A.}~\bibnamefont {Zettl}},\ }\href@noop {} {\bibfield  {journal} {\bibinfo
  {journal} {Phys. Rev. Lett.}\ }\textbf {\bibinfo {volume} {124}},\ \bibinfo
  {pages} {206403} (\bibinfo {year} {2020})}\BibitemShut {NoStop}%
\bibitem [{\citenamefont {Sancho}\ \emph {et~al.}(1985)\citenamefont {Sancho},
  \citenamefont {Sancho}, \citenamefont {Sancho},\ and\ \citenamefont
  {Rubio}}]{sancho1985highly}%
  \BibitemOpen
  \bibfield  {author} {\bibinfo {author} {\bibfnamefont {M.~L.}\ \bibnamefont
  {Sancho}}, \bibinfo {author} {\bibfnamefont {J.~L.}\ \bibnamefont {Sancho}},
  \bibinfo {author} {\bibfnamefont {J.~L.}\ \bibnamefont {Sancho}},\ and\
  \bibinfo {author} {\bibfnamefont {J.}~\bibnamefont {Rubio}},\ }\href@noop {}
  {\bibfield  {journal} {\bibinfo  {journal} {Journal of Physics F: Metal
  Physics}\ }\textbf {\bibinfo {volume} {15}},\ \bibinfo {pages} {851}
  (\bibinfo {year} {1985})}\BibitemShut {NoStop}%
\bibitem [{\citenamefont {Khomyakov}\ \emph {et~al.}(2005)\citenamefont
  {Khomyakov}, \citenamefont {Brocks}, \citenamefont {Karpan}, \citenamefont
  {Zwierzycki},\ and\ \citenamefont {Kelly}}]{khomyakov2005conductance}%
  \BibitemOpen
  \bibfield  {author} {\bibinfo {author} {\bibfnamefont {P.}~\bibnamefont
  {Khomyakov}}, \bibinfo {author} {\bibfnamefont {G.}~\bibnamefont {Brocks}},
  \bibinfo {author} {\bibfnamefont {V.}~\bibnamefont {Karpan}}, \bibinfo
  {author} {\bibfnamefont {M.}~\bibnamefont {Zwierzycki}},\ and\ \bibinfo
  {author} {\bibfnamefont {P.~J.}\ \bibnamefont {Kelly}},\ }\href@noop {}
  {\bibfield  {journal} {\bibinfo  {journal} {Physical Review B—Condensed
  Matter and Materials Physics}\ }\textbf {\bibinfo {volume} {72}},\ \bibinfo
  {pages} {035450} (\bibinfo {year} {2005})}\BibitemShut {NoStop}%
\bibitem [{\citenamefont {Ong}\ and\ \citenamefont
  {Zhang}(2015)}]{ong2015efficient}%
  \BibitemOpen
  \bibfield  {author} {\bibinfo {author} {\bibfnamefont {Z.-Y.}\ \bibnamefont
  {Ong}}\ and\ \bibinfo {author} {\bibfnamefont {G.}~\bibnamefont {Zhang}},\
  }\href@noop {} {\bibfield  {journal} {\bibinfo  {journal} {Physical Review
  B}\ }\textbf {\bibinfo {volume} {91}},\ \bibinfo {pages} {174302} (\bibinfo
  {year} {2015})}\BibitemShut {NoStop}%
\end{thebibliography}%

\clearpage\clearpage
\onecolumngrid
\setcounter{equation}{0}
\setcounter{figure}{0}
\setcounter{section}{0}
\setcounter{page}{1}

\renewcommand\theequation{S\arabic{equation}}
\renewcommand\thefigure{S\arabic{figure}}
\renewcommand\thepage{S\arabic{page}}
\renewcommand\thesection{S-\Roman{section}}

\begin{center}
	\fontsize{16}{30} \selectfont\sffamily{\supplement}\\
	\fontsize{20}{30} \selectfont\sffamily\textbf{\baslik}
\end{center}


\section{Green's Function Methodology}
\label{sec:supp_method}
The Green's function formalism is used for calculating the transmission amplitudes and  for carrying out further analysis.~\cite{datta_book_1997,ryndyk_book_2015} 
The transmission function $T(E)$ is given by
\begin{eqnarray}
	\trans(E) = \text{Tr} \left[ \Gamma_L \mathcal{G}_C \Gamma_R \mathcal{G}_C^\dagger \right],
\end{eqnarray}	
where $\mathcal{G}_C = \left[ E + i0^+ - H_C - \Sigma_L - \Sigma_R \right]^{-1}$ denotes the Green's function of the device region. In this equation, $E$ represents the energy of the system, and $0^+$ is an infinitesimally small positive number ensuring the proper analytic behavior of the Green's function. The device region is coupled to semi-infinite reservoirs on both sides, and they are assumed to be free from any scattering processes. The matrices $\Gamma_{L/R} = -2 \, \text{Im} \, \Sigma_{L/R}$ characterize the broadening of the quantum states, with $\Sigma_{L/R}$ representing the self-energy terms arising from the interaction between the central region and the reservoirs.
Transmission is computed for various disorder configurations, and ensemble averages are subsequently obtained from a sufficiently large number of realizations (at least 1000 in our case). To expedite the computational process, numerically exact decimation techniques are employed, which significantly enhance the efficiency of the calculations.~\cite{sancho1985highly}

The anomalous behaviors observed in the systems can be attributed to the realization of different transport regimes across various channels. In order to compute the individual contributions of each channel to the overall transmission, we apply the mode-matching method. This technique specifically utilizes the periodicity of the reservoirs. The transmission amplitude from the $n^{\text{th}}$ channel of the left reservoir to the $m^{\text{th}}$ channel in the right reservoir is represented as the $(n,m)$-th element of the $t$-matrix, which is defined as~\cite{khomyakov2005conductance,ong2015efficient}
\begin{eqnarray}
	t = i \sqrt{Vg_L(+)} U_L^{-1}(+) \mathcal{G}_{LR} U_R^{-1}(-)^\dagger \sqrt{Vg_R(+)}.
\end{eqnarray}
In this expression, $U_R(+)$ and $U_L(-)$ are the normalized eigenstates of the Bloch matrices for the right and left reservoirs, respectively, while $Vg_L(+)$ and $Vg_R(+)$ represent the group velocities in the respective directions. The symbols $+$ and $-$ correspond to the left- and right-going modes, respectively. It is important to note that, even though the transmission matrix elements indexed by the reservoir degrees of freedom, it still encodes information about the central region. The Green function $\mathcal{G}_{LR} = g_L H_{LC} \mathcal{G} H_{CR} g_R $ establishes a connection between the reservoir modes through the central device region, where $g_{L/R} = [E + i0^+ - H_{L/R}]^{-1}$ is the free Green's function for the left/right reservoirs.

The transmission probability from the \( m^{\text{th}} \) mode on the left to the \( n^{\text{th}} \) mode on the right is determined by taking the square modulus of the matrix element \( |t_{mn}|^2 \). By summing over \( n \), one can obtain the total transmission probability for the \( m^{\text{th}} \) mode, denoted as \( \trans^L_m = \sum_{n \in R} |t_{mn}|^2 \). Consequently, the total transmission amplitude for the system is given by
\begin{eqnarray}
	\trans = \sum_{m \in L} \trans_m^L = \text{Tr} \left[ tt^\dagger \right].
\end{eqnarray}
The eigenvalues of the matrix \( tt^\dagger \) correspond to the transmission eigenvalues, which quantify the likelihood of transport through the system. The magnitudes of these eigenvalues lie within the interval [0,1] and are highly sensitive to the disorder configuration, revealing whether a given channel is open (transmitting) or closed (non-transmitting).

\clearpage
\section{Transmission Across Edge Disordered Hexagonal Lattices}

\begin{figure}[hb!]
	\includegraphics[width=.85\textwidth]{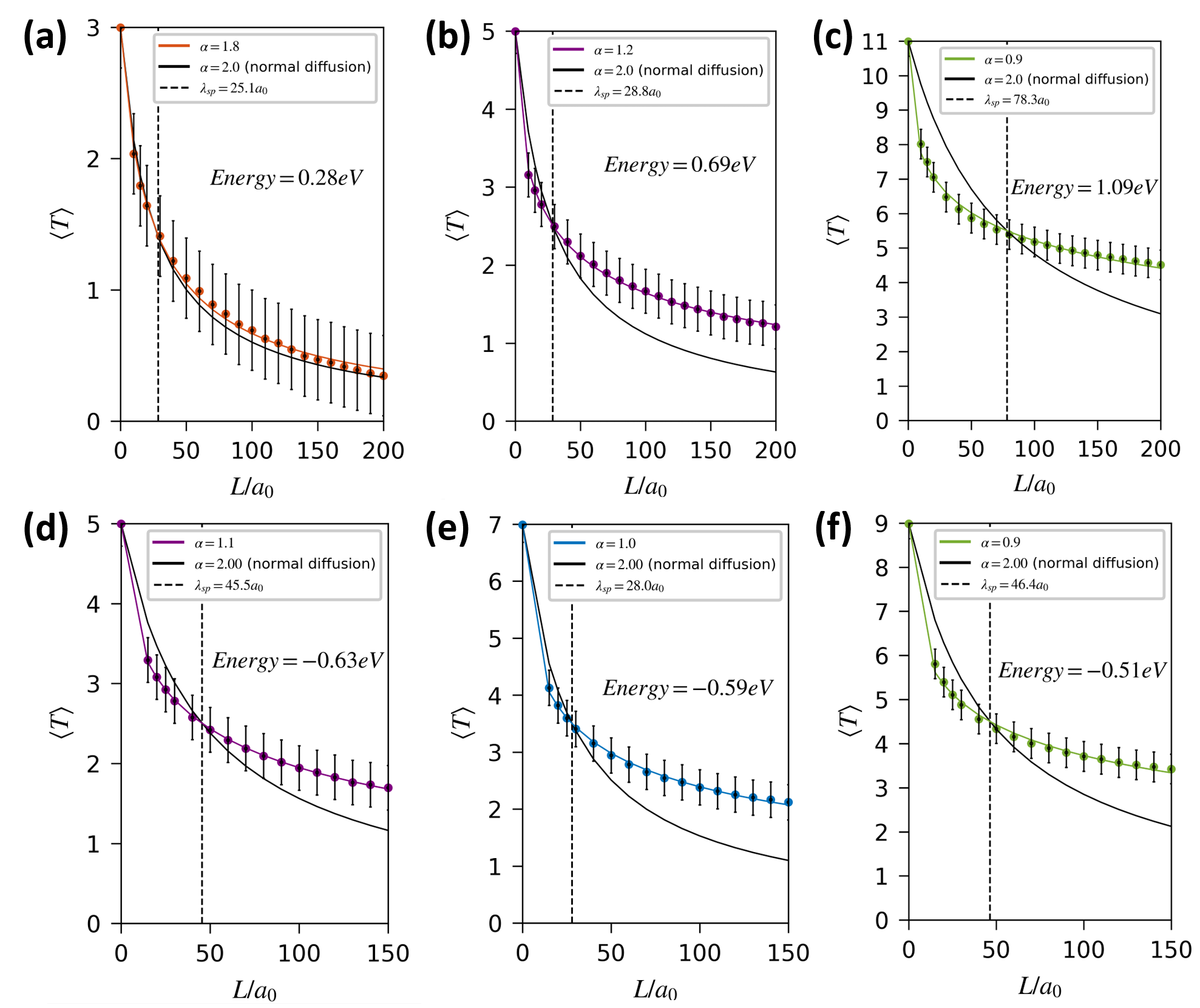}
	\caption{
		The average transmission versus system length for $20\%$ edge vacancy averaged over 1500 different disorder configurations for GNR at (a)~$E =0.28$~eV (b)~$E = 0.69$~eV (c)~$E = 1.09$~eV, and  for QNR at (d)~$E =-0.63$~eV, (e)~$E =-0.59$~eV, (f)~$E =-0.51$~eV. 
		The spread in the numerical data is calculated by using standard deviation.}
	\label{fig:supp_TransLength_Ribbon}
\end{figure}

In this part of the \supplement, the length dependent analysis of average transmission values are extended for quasi-1D grapehne and quartic nanoribbon structures.
The systems are GNR (a-c) and QNR (d-f) with $20\%$ edge vacancies.
Fig.~\ref{fig:supp_TransLength_Ribbon} shows the average transmission as a function of system length. 
The maximum values of the vertical axes  indicate the total number of channels  at that energy. 
The average transmission values are fitted to the generalized diffusion equation, and compared to the normal diffusion curve for the same $\lsp$.
Diffusion exponent values ranging between $\alpha=0.9$ to $1.8$ are observed.

For GNR at $E=0.28$~eV, Fig.~\ref{fig:supp_TransLength_Ribbon}(a), $\alpha$ is close to 2 and the spread in large. As a result, it becomes the anomalous diffusion data is not easily distinguishable from the normal diffusion. But in other cases anomalous behavior is easily distinguished from the normal diffusion, especially for $L>\lsp$ the long tail of the anomalous diffusion is identified.

\clearpage

\section{Anomalous Diffusion in Edge Disordered Square Lattice}
\label{sec:supp_square_lattice}
\begin{figure}[h]
	\includegraphics[width=1\textwidth]{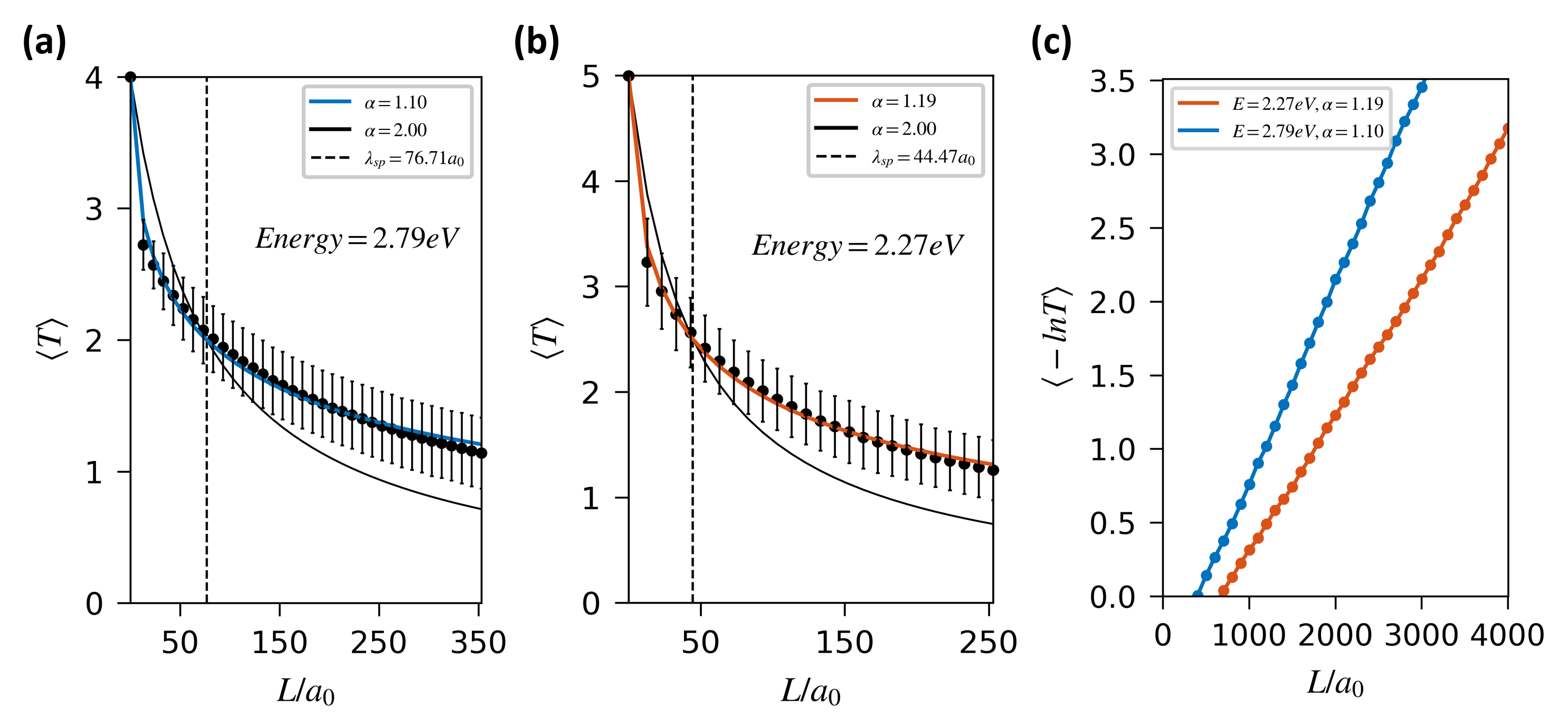}
	\caption{
		Length dependence of ensemble averaged transmission for ribbons of square lattice with edge vacancies. A $20\%$ defect density is realized for over 1000 different disorder configurations for a system width of  $10\,a_0$ at $E =2.79$~eV (a), and $E =2.27$~eV~(b). In (c), a comparison of the geometric average of the transmissions with increasing system length is given for at $E =2.79$~eV and $E =2.27$~eV.}
	\label{fig:supp_Square_TransLength}
\end{figure}

The square lattice is also investigated as a model system. 
The length dependence of transmission is investigated for two energy values.
Fig.~\ref{fig:supp_Square_TransLength}(a-b) displays the anomalous diffusion behavior with $\alpha=1.10$, and $\alpha=1.19$, respectively.
For longer systems, the linear scaling of $\langle\ln\,T\rangle$ indicates normal diffusion, see Fig.~\ref{fig:supp_Square_TransLength}(c).

\begin{figure}[b]
	\includegraphics[width=1.\textwidth]{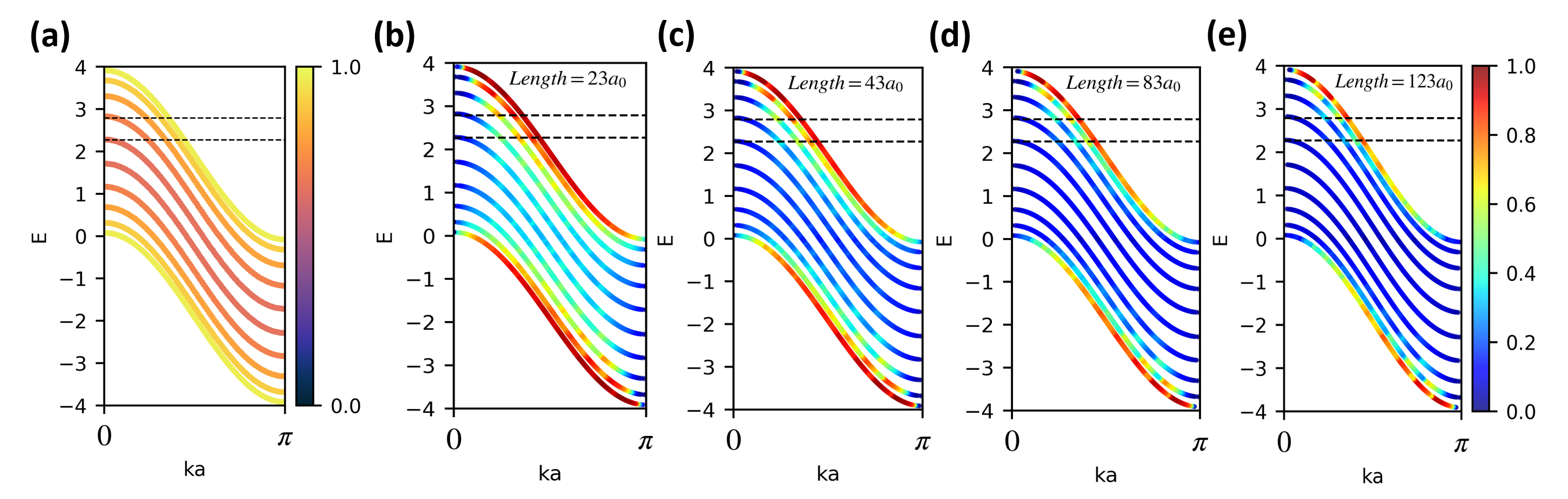}
	\caption{(a) The contribution of the edge and bulk states on the energy band diagram for the square lattice having a width of $10\,a_0$ under $20\%$ edge defect density. The band-resolved transmission probabilities for the same system having lengths of 23$\,a_0$, 43$\,a_0$, 83$\,a_0$, and 123$\,a_0$ are plotted with averaging over ensembles with 1000 different disorder configurations (b-e).
	}
	\label{fig:supp_Square_BandResolved}
\end{figure}

In Fig.~\ref{fig:supp_Square_BandResolved}(a), the edge versus bulk character of the bands are shown, where yellow color indicates bulk character and darker color stands for edge localized states.
The band resolved transmission values are plotted for the entire spectrum in Fig.~\ref{fig:supp_Square_BandResolved}(b-e) for different system lengths.
One observes that bulk-like states stay quasi ballistic for longer distances, whereas the edge states localize at much shorter distances.

\clearpage

\section{Channel-Resolved Transmission Probability Distributions}
\label{sec:supp_prob_dist}

Anomalous diffusion across multichannel systems is revealed through an analysis of channel-resolved transmission spectra.
Those analyses are given in Fig.~\ref{fig:fig_2channel_main} and Fig.~\ref{fig:ribbon_main}.
Here, we include further details on channel-resolved transmission probabilities and their distributions.

\subsection{Single Channel Transmission Probability Distribution}
\label{sec:supp_1channel_prob_dist}

\begin{figure}[b]
	\includegraphics[width=.66\textwidth]{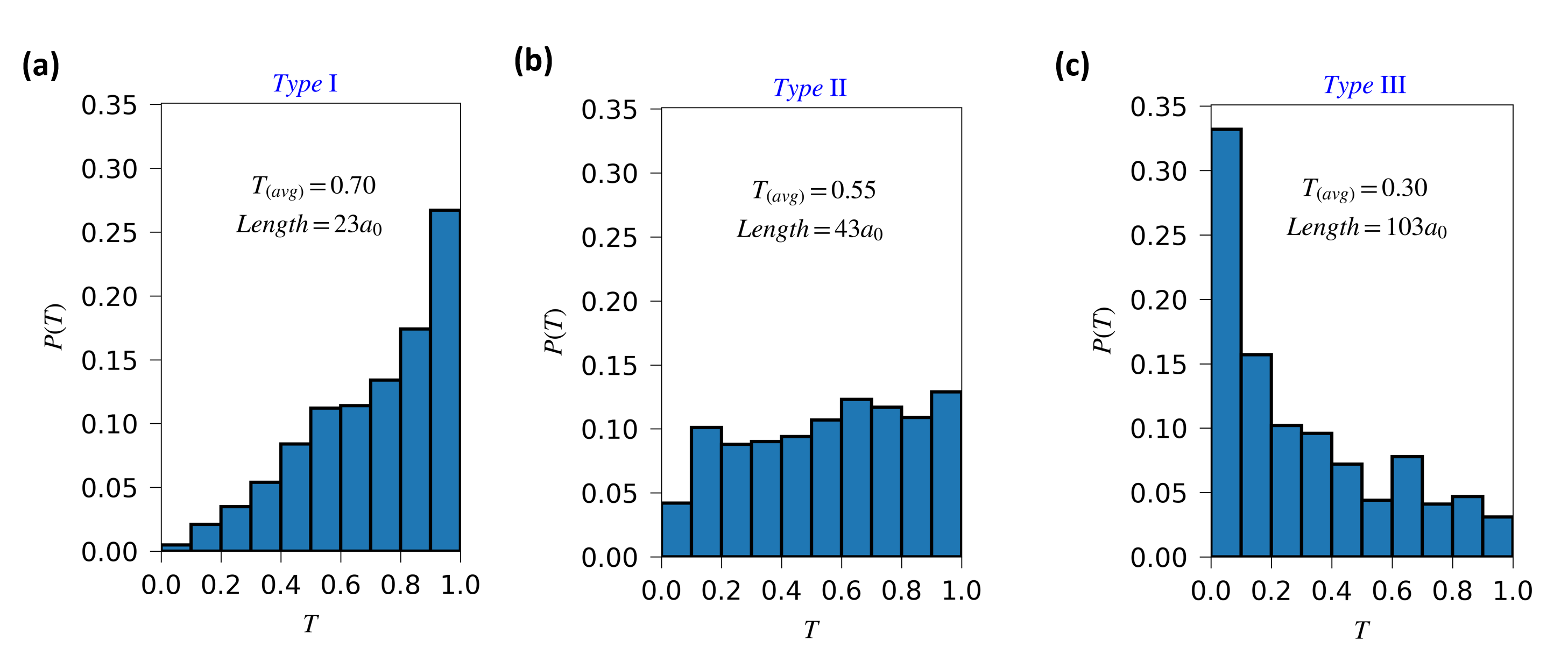}
	\caption{The possible transmission probability distributions for single-channel systems where $t_{\parallel}=1$ and $W =t_{\parallel}$. The distributions obtained over 1000 different disorder configurations at the lengths $L = 23a_0$, $L = 43a_0$, and $L =103 a_0$. 
	}
	\label{fig:supp_TransProb_SingleChannel}
\end{figure}

Prior to the analysis of multichannel systems, we first investigate transmission probability distribution for a single channel to serve as a reference point.
The hopping parameter and the strength of Anderson disorder are chosen as   $t_{\parallel}=1$ and $W =t_{\parallel}$, respectively. 
Length dependent transmission probabilities are computed for 1000 realizations.
The mean-free-path and $\lsp$ are the same and equal to $49.6\,a_0$. 
Fig.~\ref{fig:supp_TransProb_SingleChannel} shows transmission probability distributions for system lengths of $ L = 23a_0$, $L = 43a_0$ and $L = 103a_0$.
These distributions clearly mark the nearly-ballistic (a), diffusion (b), and localization (c) regimes.

\begin{figure}[b]
	\includegraphics[width=1.\textwidth]{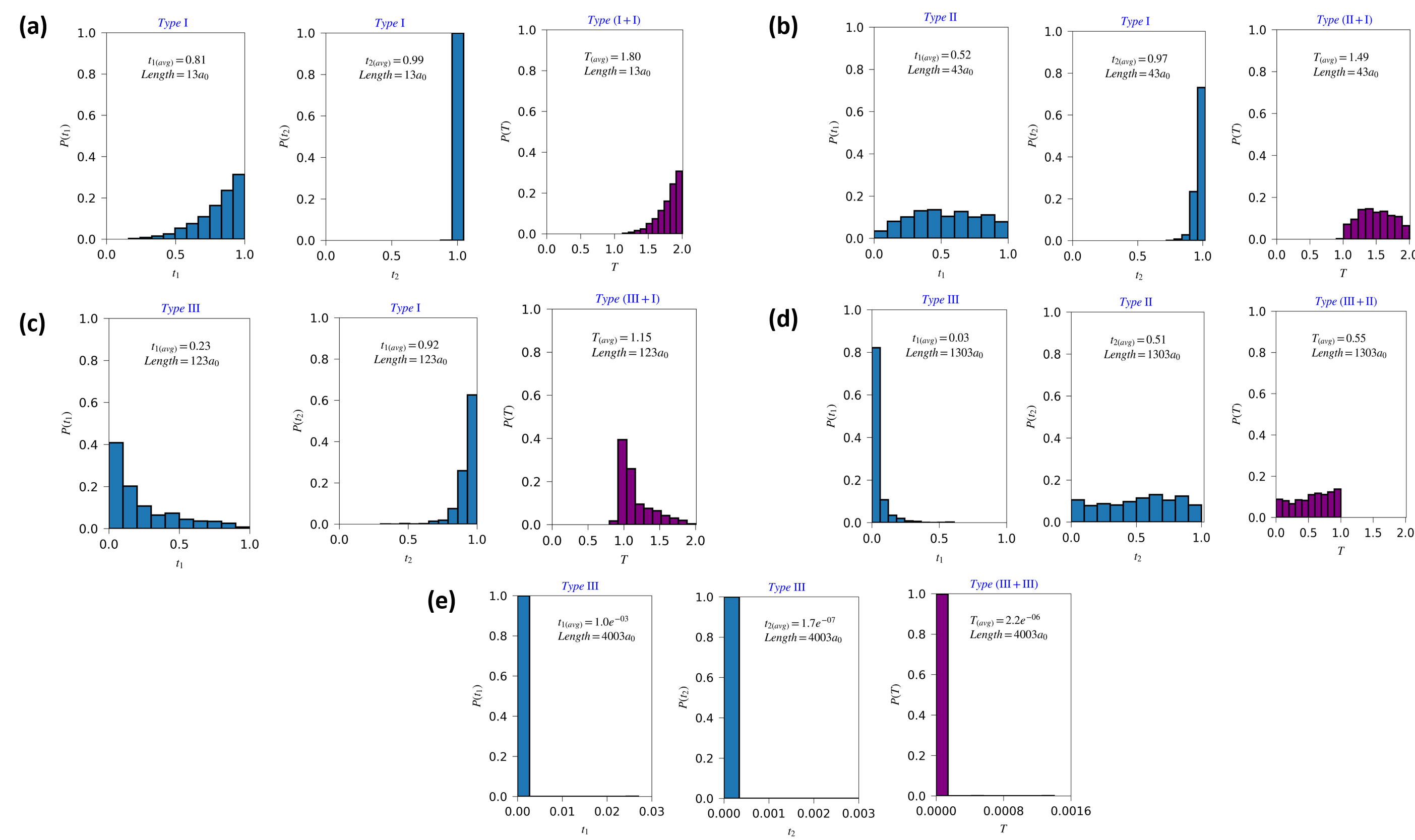}
	\caption{
		The transmission probability distributions at different lengths for $(a)L = 13a_0$ (b)$L = 43a_0$ (c)$L = 123a_0$ (d)$L = 1303a_0$ (e) $L = 4003a_0$ where the spread length of the system is $207.43 a_0$ and $t_{\perp}/t_{\parallel} = 1/20$. Individual distributions indicated with  $P(t_1)$ and $P(t_2)$, for Channel-1 and Channel-2, obtained from the mode-matching method while the transmission distribution of the total system, $P(T)$, is calculated using Green’s function method at the same lengths.}	
	\label{fig:supp_TransProb_TwoChannel}
\end{figure}

\begin{figure}[t]
	\includegraphics[width=1.\textwidth]{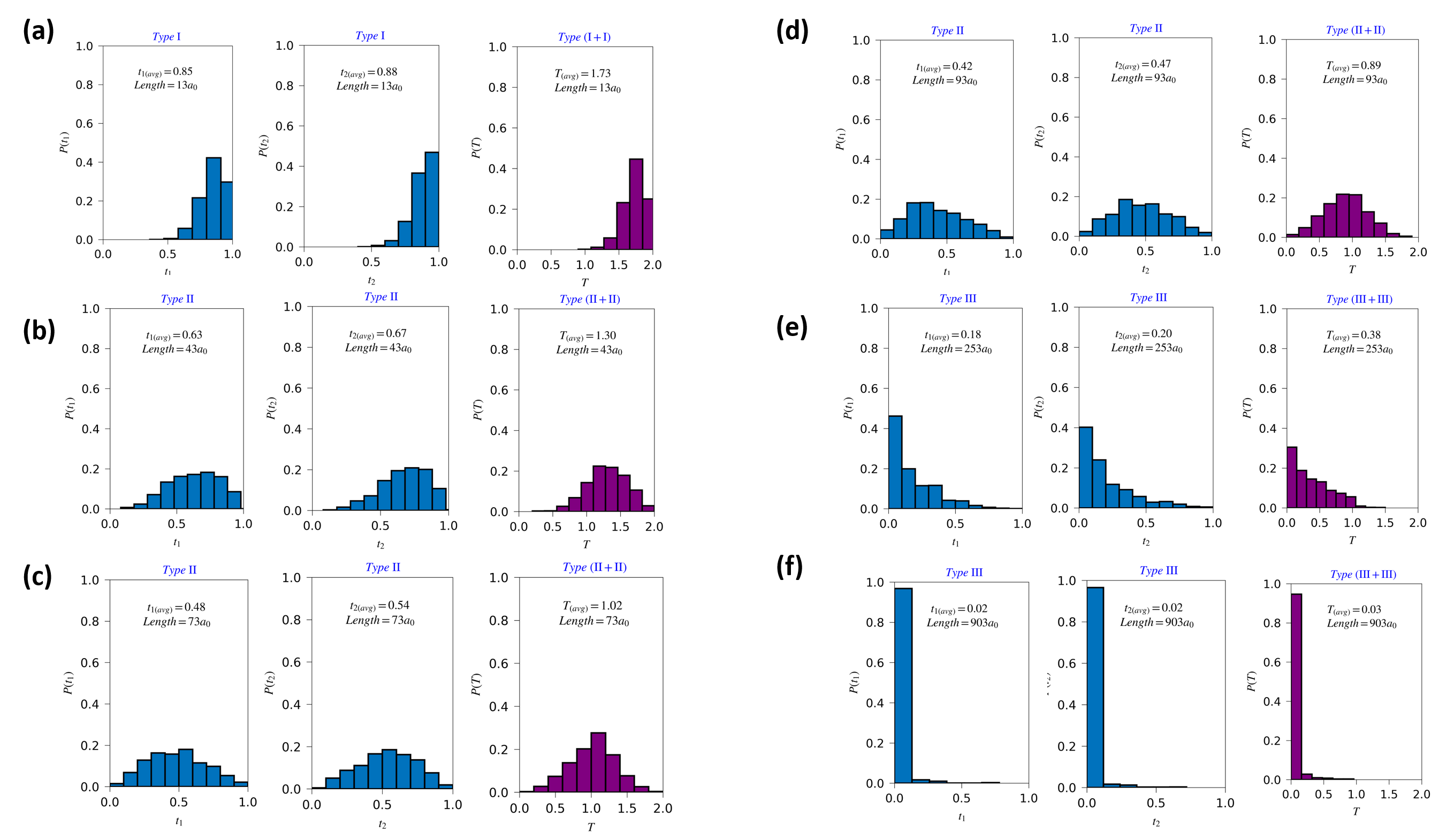}
	\caption{
		Channel-resolved transmission probability distributions in the case of strong inter-channel coupling are in stark contrast with that of weak inter-channel coupling case presented in Fig.~\ref{fig:supp_TransProb_TwoChannel}.
	}
	\label{fig:supp_BandResolved_TwoChannel_StrongInterChannelCoupling}
\end{figure}

\subsection{Transmission Probability Distributions: Two-Channel Model}
\label{sec:supp_2channel_prob_dist}

Next, we analyze transmission probability distributions for the two-channel model. The analysis is performed for five systems lengths, namely, 13$\,a_0$, 43$\,a_0$, 123$\,a_0$, 1303$\,a_0$, and 4003$\,a_0$.
The corresponding probability distributions are plotted in Fig.~\ref{fig:supp_TransProb_TwoChannel}(a-e).
In the first two panels, channel-resolved probability distributions are plotted in blue for Channel-1 and Channel-2, respectively. The third panel shows the  unresolved distribution (purple).

Depending on the system length, channels represent ballistic, diffusion or localization characters that were observed for a single channel in Fig.~\ref{fig:supp_TransProb_SingleChannel}.
For example, for $L=13\,a_0$, both channels posses ballistic character (Fig.~\ref{fig:supp_TransProb_TwoChannel}(a)). 

For $L=43\,a_0$, the unresolved probabilities showcase an unusual distribution in the third panel of Fig.~\ref{fig:supp_TransProb_TwoChannel}(b). Probable transmission values are distributed almost evenly, but only within the range between 1 and 2. This does not resemble any of the distributions for a single channel system, or the two-channel system with $t_\perp/t_\parallel=1$. 
The channel-resolved plots unravel the underlying reason.
Namely, Channel-1 possesses a diffusion distribution, while Channel-2 still preserves its ballistic nature.
As a result, total transmission is almost always larger than or equal to 1, and the distribution of values between 1 and 2 are similar to that of a single diffusive channel.

\begin{figure}[t]
	\includegraphics[width=1.\textwidth]{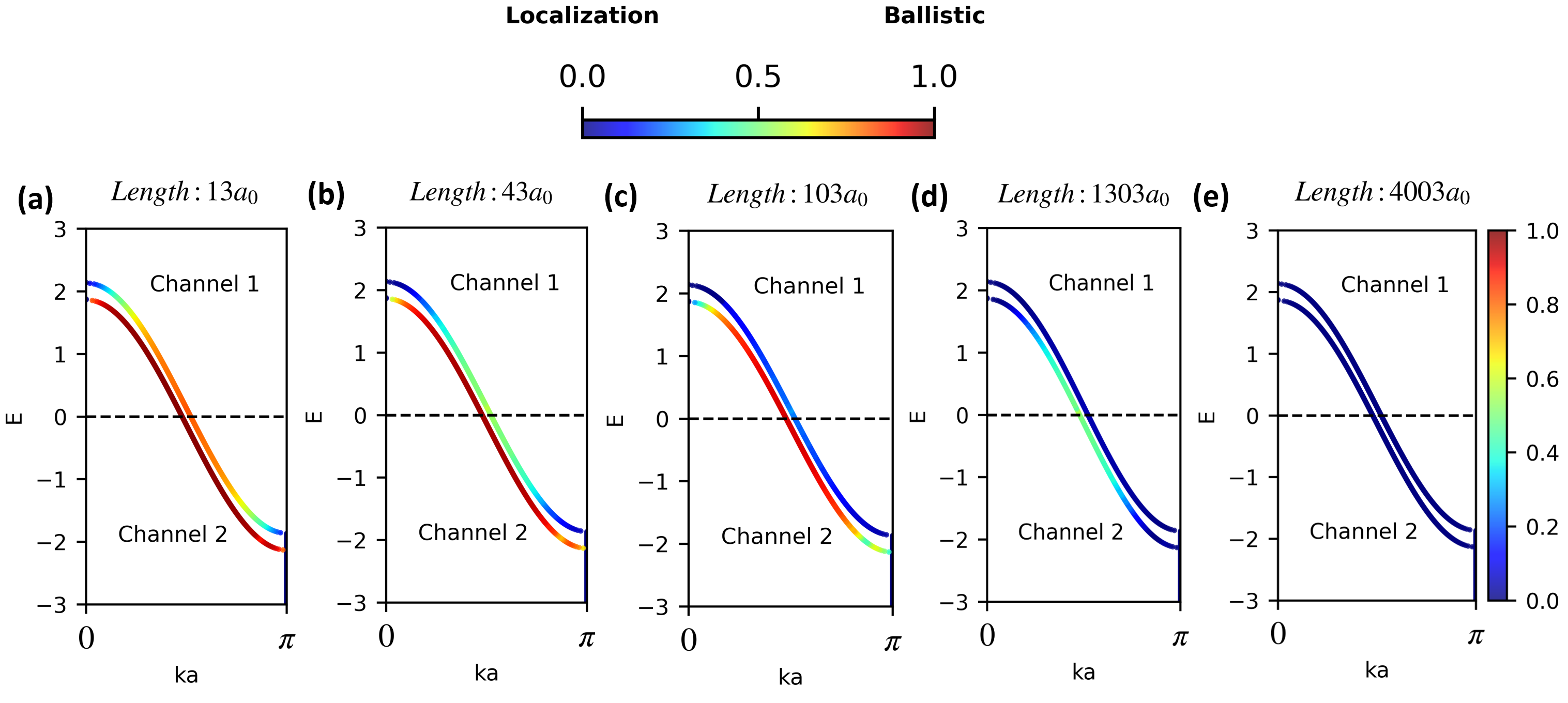}
	\caption{The band resolved probabilities for the two-channel model with increasing system lengths where $t_{\perp}/t_{\parallel} = 1/20$. The spread length of the system ($\lsp$) obtained as $207.43\,a_0$}	
	\label{fig:supp_BandResolved_TwoChannel}
\end{figure}

As the system gets longer, specifically for $L=123\,a_0$, Channel-1 enters the localization regime, but Channel-2 still has the ballistic character as shown in Fig.~\ref{fig:supp_TransProb_TwoChannel}(c).
The unresolved distribution has a pronounced peak around $T=1$, and it is markedly asymmetric. The distribution to the left of the peak is determined mainly by the ballistic channel, therefore the tail is short. The right side of the peak shows a broader distribution because it is mainly determined by Channel-1, which is in the localization regime.

At $L=1303\,a_0$, Channel-2 acts diffusive, while the localization is enhanced for Channel-1, and the unresolved transmission distribution is more or less even within $[0,1]$ range, Fig.~\ref{fig:supp_TransProb_TwoChannel}(d).
Lastly, at even longer distances, the only contribution to transmission is through localized states in both channels as shown in Fig.~\ref{fig:supp_TransProb_TwoChannel}(e) for $L=4003a_0.$

Using the three transmission distribution profiles of the single channel system (\textit{i.e.} ballistic, diffusion, localization), one can make six different pairs of distribution profiles. In the above analysis we have shown five of them, simply because diffusion-diffusion pair was not accessible within the chosen parameters. 
Three of the cases involve pairings of dissimilar transmission regimes: ballistic-diffusion, ballistic-localization, diffusion-localization, which are associated with the anomalous transport behavior.

We note that the observed effects are not limited to specific energy windows but could be widely observable. Fig.~\ref{fig:supp_BandResolved_TwoChannel} shows average transmission values for the same systems as in Fig.~\ref{fig:supp_TransProb_TwoChannel} for all possible energies. Evidently, the pairings of dissimilar transport regimes, and anomalous diffusion emerges in the predominant part of the spectrum.

\clearpage
\section{Transmission Probability Distribution in Edge-Disordered Hexagonal Lattices}
\label{sec:supp_hexagonal_prob_dist}

\begin{figure}[h]
	\includegraphics[width=.9\textwidth]{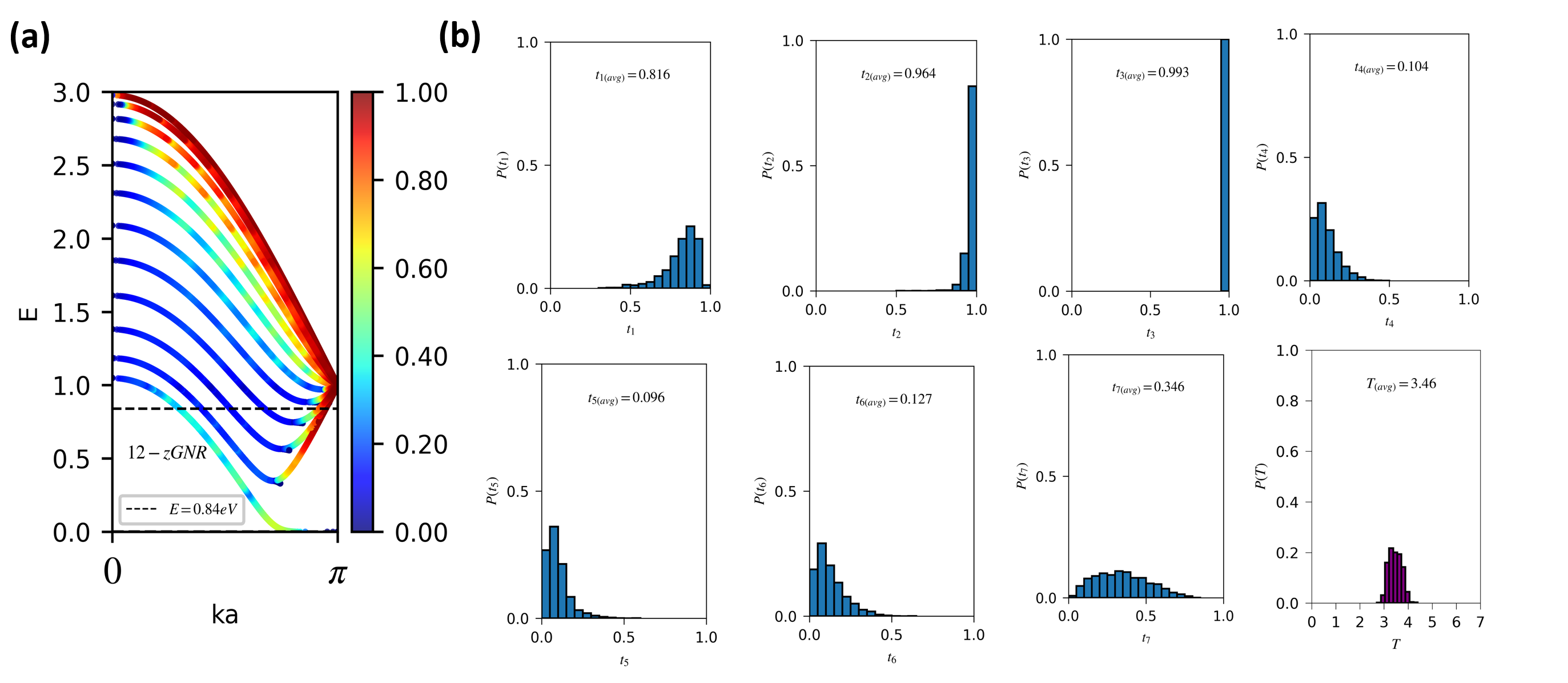}
	\caption{The band-resolved probabilities of GNR
		at $L = 40\,a_0$ where the system has $20\%$ defect density. At 0.84~eV, $\lsp$ obtained as 39.3~$a_0$ and the distribution of the transmission probabilities for each channel calculated over 1500 different disorder configurations.}
	\label{fig:supp_TransProb_GNR}
\end{figure}

\begin{figure}[h]
	\includegraphics[width=.9\textwidth]{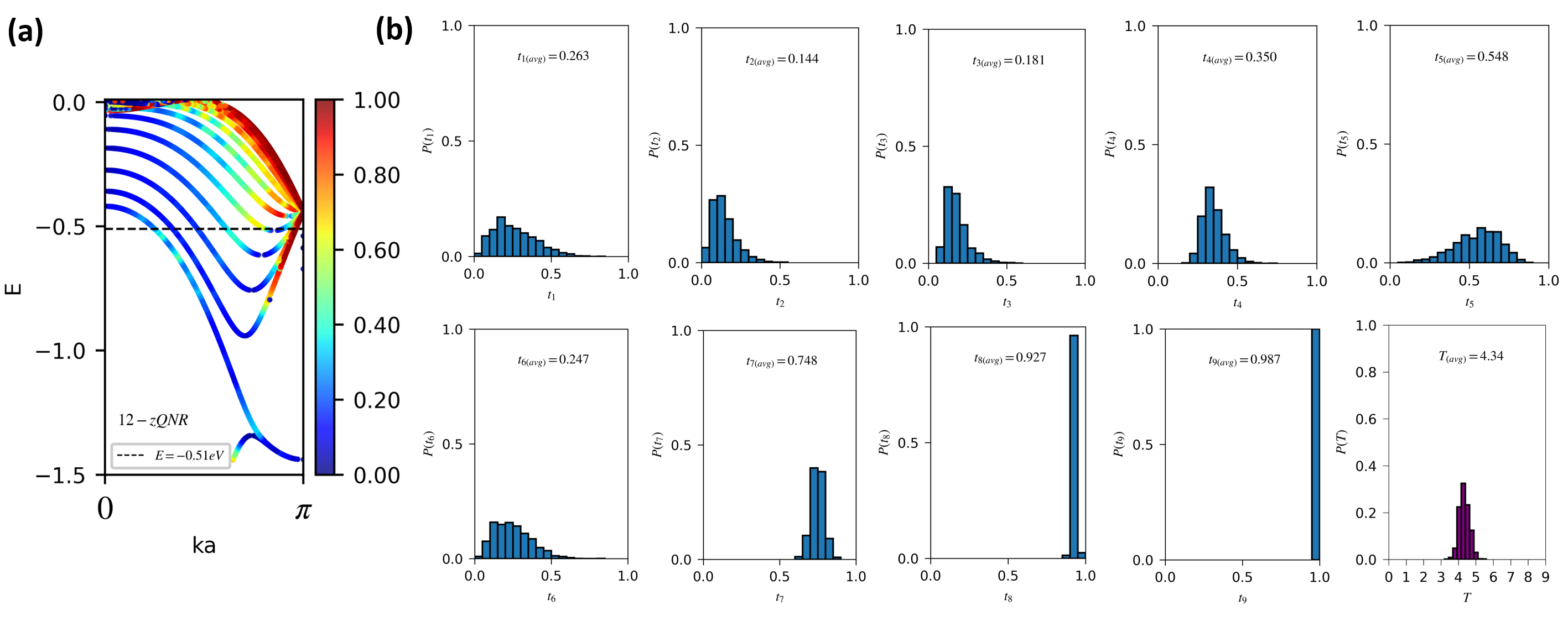}
	\caption{
		The band-resolved probabilities of GNR
		at $L = 50\,a_0$ where the system has $20\%$ defect density. At -0.51~eV, $\lsp$ is $46.35\,a_0$. Transmission probability distributions for each channel are calculated using 1500 different disorder configurations.}
	\label{fig:supp_TransProb_QNR}
\end{figure}

	By following the same steps as the calculations of the two-channel model, we obtain the channel-resolved transmission probabilities for hexagonal nanoribbons using mode-matching analysis.
	Fig.~\ref{fig:supp_TransProb_GNR} shows the transmission distributions of each channel at 0.84 eV averaged over 1500 different disorder configurations for GNR. These calculations were performed at $L =80\,a_0$, nearly equal to the spread length at that energy $(\lsp= 39.34\,a_0)$. The energy band diagram shows that the system consists of seven different channels. According to the mode matching analysis results, the average transmission probabilities vary for these channels from 0.10 to 0.99. 
	This enormously wide range shows that different channels can be found in different transport regimes simultaneously, which is the origin of the anomalous diffusion in our systems.
	Similarly, Fig.~\ref{fig:supp_TransProb_QNR} shows the transmission distributions of each channel at -0.51~eV  and $20\%$ defect density over 1500 different disorder configurations for QNR. According to the energy band diagram, at that energy, the quartic nanoribbon has nine different channels, and each of them has different transmission probabilities whose average ranges from 0.14 to 0.98.


\clearpage

\begin{figure}[b]
	\includegraphics[width=0.5\textwidth]
	{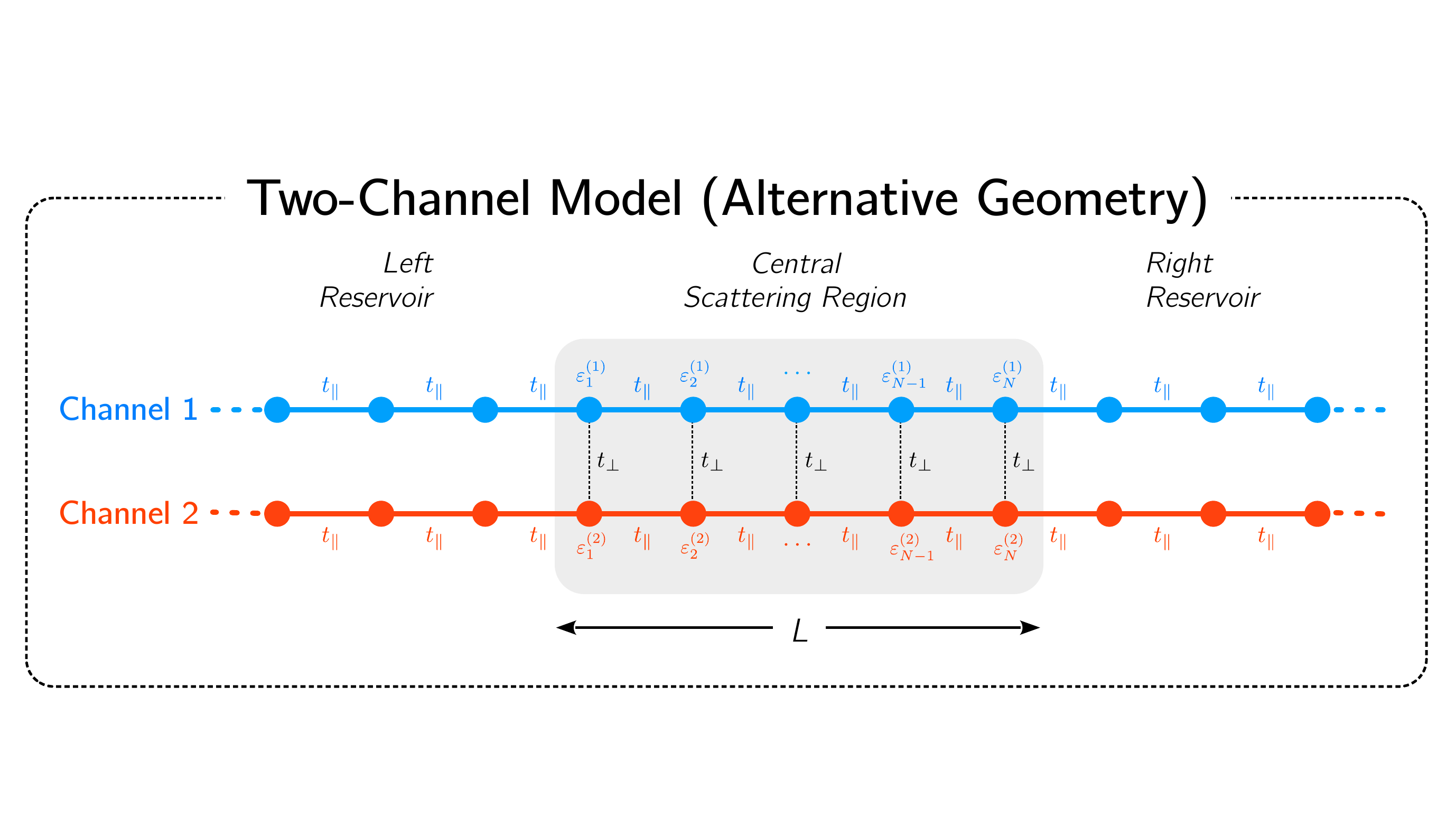}
	\includegraphics[width=0.5\textwidth]{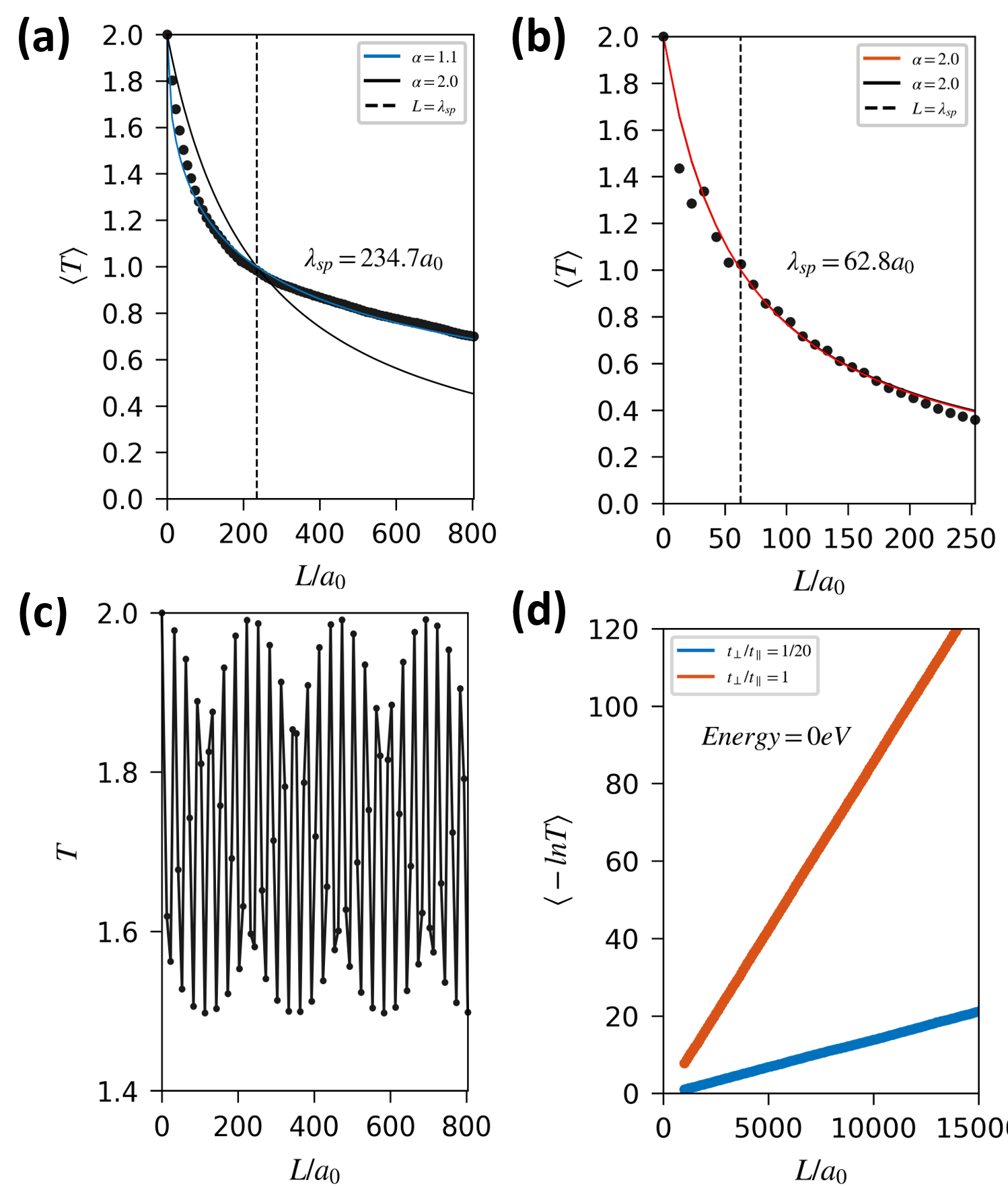}
	\caption{
		\textbf{An alternative geometry for the two-channel model} can be chosen with  $t_\perp=0$ in the reservoirs.
		Weak (a) and strong (b) interchannel coupling cases are considered.
		Length dependent transmission averaged over 1000 configurations is plotted for $E=0$~eV. 
		}
	\label{fig:SUPP_2channel_alternative}
\end{figure}

\section{An Alternative Geometry for the Two-Channel Model}
\label{sec:Supporitng_AlternativeGeometry}

The two-channel model can also be designed such that the channels are decoupled in the reservoir regions as shown in the top panel of Fig.~\ref{fig:SUPP_2channel_alternative}.
The vanishing $t_\perp$ in the reservoirs prevents hybridization of the channels.
Compared to the ladder geometry shown in Fig.~\ref{fig:fig_2channel_main}, the channels are distinguishable even when $t_\perp/t_\parallel$ is not small.

Using the same parameters with the ladder geometry, it is also possible to observe the normal and anomalous diffusion in this alternative geometry.
In Fig.~\ref{fig:SUPP_2channel_alternative}(a), one observes that $\alpha=1.1$ in the weak coupling case, whereas $\alpha=2$ for strong coupling.
We note that, for the strong coupling case, one observes fluctuations in length dependent transmission, especially for  $L<\lsp$.
These are mainly results of Fabry-Pérot oscillations due to interference within the scattering region. The oscillations are demonstrated in Fig.~\ref{fig:SUPP_2channel_alternative}(c), where transmission amplitude is plotted for a disorder-free system.
Such oscillations are not observed in the ladder geometry because the Hamiltonian is the same inside and outside the scattering region, except for the disorder terms.
In Fig.~\ref{fig:SUPP_2channel_alternative}~(d) we show that $-\langle\ln\,T\rangle$ is linear in $L$, and therefore normal localization is observed in the alternative geometry, just like in the ladder geometry.

\clearpage
\section{Four-probe Resistance}
\label{sec:Supp_FourProbe}

\begin{figure}[h]
	\includegraphics[width=0.66\textwidth]{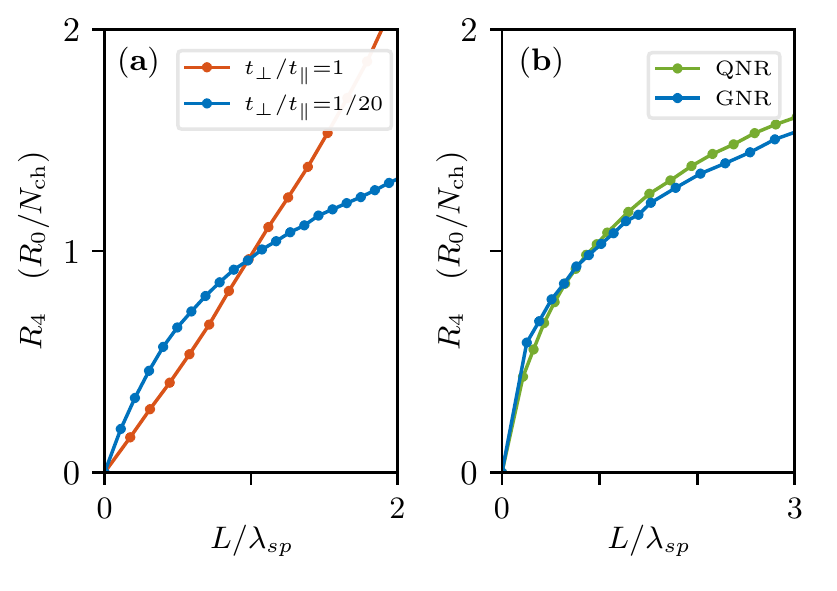}
	\caption{
	\textbf{Four-probe resistance} plotted as a function of $L/\lsp$ and in units of $R_0/\Nch$.
			The two-channel model produces a linear dependence on $L$ when $t_\parallel/t_\perp=1$,
			whereas it is sublinear $t_\parallel/t_\perp=1/20$ displaying the onset of CAAD when the channels are asymmetric and weakly coupled (a).
			Sublinear behavior in $R_4$ is intrinsic in GNR and QNR when edge disorder is present (b).
	}
	\label{fig:Supp_FourProbe}
\end{figure}

The four-probe resistance, $R_4$, depends on the system length, $L$, and the spread-length, $\lsp$ as
\begin{align}
	R_4 = \frac{R_0}{\Nch} \left( \frac{L}{\lsp} \right)^{\alpha/2}.
\end{align}
Here, $R_0$ is the quantum of resistance, $\Nch$ is the number of channels, and $\alpha$ is the diffusion exponent.
In conventional conductors, $R_0$ increases linearly with $L$, whereas in superdiffusion $R_4$ is sublinear. 
In Fig.~\ref{sec:Supp_FourProbe}(a), normal and anomalous diffusion are shown within the two-cahnnel-model.
The sublinear increase of $R_4$ with $L$ is shown for hexagonal lattices in Fig.~\ref{fig:Supp_FourProbe}(b).

\end{document}